\documentclass[12pt,a4paper]{article}
\usepackage{xcolor}
\usepackage{jheppub}
\usepackage{epstopdf}
\usepackage{graphicx}
\usepackage{epsfig}
\usepackage{dcolumn}  
\usepackage{bm}    
\usepackage{amssymb} 
\usepackage{amsmath,bm}
\usepackage{amsfonts}    
\usepackage{slashed}  
\usepackage{mathtools}
\usepackage[mathscr]{euscript}
\usepackage{epsfig}
\title{Heterotic string on the CHL orbifold of K3}

\author{Shouvik Datta$\null^{a, b}$,} 
\author{Justin R.~David${}^{b}$} 
\author{and Dieter L\"{u}st${}^{c, d}$ } 

\affiliation{ ${}^{a}$ Institut f\"{u}r Theoretische Physik, ETH Z\"{u}rich, \\
CH-8093 Z\"{u}rich, Switzerland.\vspace{.1cm} \\
${}^{b}$  Centre for High Energy Physics,
Indian Institute of Science,\\ C.V. Raman Avenue, Bangalore 560012, India. \vspace{.1cm} \\  
${}^{c }$ 
Arnold Sommerfeld Center for Theoretical Physics,  \\
Theresienstrasse 37, 80333 M\"{u}nchen, Germany.  \vspace{.1cm}\\  
${}^{d }$Max-Planck-Institut f\"{u}r Physik \\
F\"{o}hringer Ring 6, 80805 M\"{u}nchen, Germany. 
}

\emailAdd{shouvik@itp.phys.ethz.ch, justin@cts.iisc.ernet.in, dieter.luest@lmu.de}

\preprint{MPP-2015-255,
LMU-ASC 66/15}

\abstract{ We study ${\cal N}=2$ compactifications of heterotic string theory on
the CHL orbifold  $(K3\times T^2)/\mathbb{Z}_N$ with $N=  2, 3, 5, 7$. 
$\mathbb{Z}_N$ acts as an involution on $K3$ together with a shift of $1/N$ along one
of the circles of $T^2$. 
These compactifications generalize 
the example of the heterotic string on $K3\times T^2$ studied in the 
context of dualities in ${\cal N}=2$ string theories. 
We evaluate the new supersymmetric index for these theories   and show that 
their expansion can written in terms of the McKay-Thompson series associated with the 
$\mathbb{Z}_N$ involution embedded in  the Mathieu group $M_{24}$.  
We then evaluate the difference in 
one-loop  threshold corrections to the non-Abelian gauge couplings with Wilson lines and show 
that their moduli dependence  is captured  by Siegel modular forms  related to dyon partition functions 
of ${\cal N}=4$ string theories. 
}

\begin{document}
 \maketitle

\def\nn{\nonumber}
\def\pd{\partial}
\def\Re{R\'{e}nyi }
\def\l1{{{1-loop}}}
\def\uy{u_y}
\def\ur{u_R}
\def\o{\mathcal{O}}
\def\Cl{{{cl}}}
\def\bz{{\bar{z}}}
\def\by{{\bar{y}}}
\def\bX{\bar{X}}
\def\im{{{Im}}}
\def\re{{{Re}}}
\def\cn{{{cn}}}
\def\sn{{{sn}}}
\def\dn{{{dn}}}
\def\K{\mathbf{K}}
\def\n1{\Bigg|_{n=1}}
\def\fin{{{finite}}}
\def\R{{\mathscr{R}}}
\def\one{{(1)}}
\def\zero{{(0)}}
\def\n{{(n)}}
\def\tr{{Tr}}
\def\T{\mathcal{T}}
\def\TT{\tilde{\mathcal{T}}}
\def\O{\mathcal{O}}
\def\cN{\mathcal{N}}
\def\P{\Phi}
\def\W{{\tilde{W}}}
\def\T{{\tilde{T}}}
\def\I{\ \mathbf{\mathcal{I}}}
\def\ln{\boldsymbol{\colon}\hspace{-.13cm}}
\def\rn{\hspace{-.08cm}\boldsymbol{\colon}\hspace{-.05cm}}
\def\see{S_{{EE}}^{(2)}}
\def\rp{\rho_P}
\def\rq{\rho_Q}
\def\xpp{x^+_P}
\def\xpq{x^+_Q}
\def\xmp{x^-_P}
\def\xmq{x^-_Q}
\def\L{\, \mathbf{\mathsf{\Lambda}}}
\def\F{\mathcal{F}}
\def\vt{\vartheta}
\def\half{\frac{1}{2}}
\def\tr{\text{Tr}}
\def\th{\theta}
\def\blue#1{\textcolor{blue}{#1}}
\def\cZ{\mathcal{Z}}
\def\Znew{\mathcal{Z}_{\text{new}}}
\def\E{\mathscr{E}}
\def\chl{{\text{CHL}}}
\def\F{\mathcal{F}}
\def\red#1{\textcolor{red}{#1}}
\def\blue#1{\textcolor{blue}{#1}}
\def\B{\mathcal{B}}

\section{Introduction}

${\cal N}=2$ compactifications of  heterotic string theory have proved to be good testing 
ground to explore duality symmetries of string theory. 
One of the main motivations to explore these  compactifications is that these vacua 
have dual realization in terms of type II compactifications on Calabi-Yau.  
Identifying dual pairs on the heterotic and type II side  enables highly non-trivial 
tests of dualities with ${\cal N}=2$ symmetry \cite{Kachru:1995wm}. 
The simplest example of such theories is the heterotic string theory compactified 
on $K3 \times T^2$.  This theory was first constructed in  $d=6$ in \cite{Green:1984bx,Walton:1987bu}. 
An important observable for the test of duality  in this theory is the dependence of the  one-loop corrections of 
gauge and gravitational coupling constants on the vector multiplet moduli of the theory. 
The moduli dependence of these threshold corrections are 
encoded in automorphic forms of the heterotic duality group 
\cite{Dixon:1990pc,Mayr:1995rx,deWit:1995zg,Harvey-Moore,deWit:1996wq,Cardoso-Curio-Lust}.

Our goal in this paper is to first consider more general compactifications of  the heterotic
string on $(K3 \times T^2)/\mathbb{Z}_N$, with $N=2, 3, 5, 7$. 
$\mathbb{Z}_N$ acts by a $1/N$ shift on one of the circles of $T^2$ together with  an  action 
on the internal CFT describing  the heterotic string theory on $K3$. 
This freely acting orbifold of  $K3\times T^2$ was  first studied on the type II side first as duals of 
CHL compactifications \cite{Chaudhuri:1995fk,Chaudhuri:1995bf}
 of the heterotic string \cite{Aspinwall:1995fw,Ferrara:1989nm,Schwarz:1995bj}.
We will call this orbifold,  the CHL orbifold of $K3$. 
These compactifications of the heterotic string on the CHL orbifold of 
$K3$ preserve  ${\cal N}=2$ supersymmetry and the number 
of vector multiplets, but reduce the the number 
of  charged and un-charged hypermultiplets in the theory. 
They also affect the vector multiplet moduli dependence of  the one-loop corrections. 
The two main aspects of these compactifications we study  in this paper are  the new supersymmetric index 
and  the gauge threshold corrections.   
We summarize the results  obtained in the next few paragraphs.

The basic quantity from which one-loop thresholds of heterotic string 
on $K3 \times T^2$  are obtained is the 
new supersymmetric index 
\cite{Antoniadis:1992sa,Antoniadis:1992rq,Cecotti:1992qh,Cecotti:1992vy,Harvey-Moore,Cardoso-Curio-Lust} which is  defined as 
\begin{equation}
 {\cal Z}_{\rm new} 
  (q, \bar q )  = \frac{1}{\eta^2(\tau)  } {\rm Tr}_{R} 
 \left(    F e^{i \pi  F } q^{L_0 - \frac{c}{24} } \bar q^{\bar L_0 -\frac{ \bar c}{24}}   \right)  \ . 
\end{equation}
The trace in the above expression is taken over the Ramond sector in  the internal CFT with central 
charges $(c, \bar c) = ( 22, 9 ) $.  Here  $F$ is the  world sheet fermion number of the right moving 
 ${\cal N}=2$ supersymmetric internal CFT. 
For the standard embedding of  the spin connection  into a $SU(2)$ of one of the $E_8$'s 
of the heterotic string,  it was shown 
\cite{Harvey-Moore,Cardoso-Curio-Lust} that this index 
decomposes as 
\begin{eqnarray}\label{fact1}
 {\cal Z}_{\rm new} (q, \bar q )  & =&  \frac{8}{ \eta^{12}} \Gamma_{2, 2} ( q, \bar q )   E_4 (q)  \times 
 \frac{E_6(q)}{\eta^{12}},    \\ \nonumber
 &=& \frac{8}{ \eta^{12}} \Gamma_{2, 2} ( q, \bar q )   E_4 (q) 
  \left[
  \frac{ \theta_2(\tau)^6 }{\eta(\tau)^6  }  Z_{K3}   ( q, -1)   + 
  q^{\frac{1}{4}}  \frac{ \theta_3(\tau)^6 }{\eta(\tau)^6  }  Z_{K3  } (  q, -q^{1/2} )  \right. 
\nonumber  \\   \label{fact}
  & & \left.  \qquad\qquad\qquad \qquad \ \ 
  -  q^{\frac{1}{4}}  \frac{ \theta_4(\tau)^6 }{\eta(\tau)^6  }  Z_{K3  } (  q, q^{1/2} )
  \right] \ .
\end{eqnarray}
Here $E_4, E_6$ refer to Eisenstein series of weight $4, 6$ respectively, 
 $Z_{K3}(q, z) $ is the elliptic genus of  the  ${\cal N}= 4$  conformal field theory of $K3$ and  
 \begin{equation}
\frac{\Gamma_{10, 2}}{\eta^{10}} = \frac{1}{\eta^{10}}  \Gamma_{2, 2}( q, \bar q) E_4 (q)  \ ,
\end{equation}
is the partition function for the second  $E_8$ lattice along with the lattice 
from $T^2$. 
In \cite{Kachru},  it was shown that due to the factorization of the new supersymmetric index 
as given in  second equation of (\ref{fact}),   
the BPS states of  the heterotic compactifications  on $K3\times T^2$  have a 
decomposition in terms of representation of the Mathieu group $M_{24}$. 
We will evaluate the new supersymmetric index for  heterotic compactifications of the CHL orbifolds of $K3$ 
and show that new supersymmetric index is given by the same form as in (\ref{fact}) 
but now with $Z_{K3}(q, z) $ replaced by  the 
twisted elliptic genus of the  CHL orbifolds of $K3$.
We will evaluate the new  supersymmetric index   explicitly for the $N=2$ CHL orbifold $(K3 \times T^2)/\mathbb{Z}_2$
and then generalize this for the other values of $N$ using results of \cite{justin1}. 
We then generalize the observation of \cite{Kachru} and  show that the BPS states 
 for  heterotic compactifications of the CHL orbifolds of $K3$  have a decomposition in terms of 
 representations of the  Mathieu group $M_{24}$.

Threshold corrections are important observables in string compactifications 
and there has been a recent revival in studying properties of these observables 
 mainly due to the work of 
\cite{Angelantonj:2011br,Angelantonj:2012gw,Angelantonj:2013eja,Angelantonj:2015rxa}.
Let us  examine the  threshold corrections evaluated in  
$K3 \times T^2$ compactifications which we will generalize in this work to CHL orbifolds of $K3$. 
For concreteness consider  the standard embedding in which the spin connection 
connection of $K3$ is equated to the gauge connection. Starting from the 
$E_8 \times E_8$ theory compactifying on $K3 \times T^2$ at generic points of 
the moduli space of $T^2$ results in $E_7 \times E_8 \times U(1)^4$. 
Let the $E_8$ which is broken  to $E_7$ be referred to $G'$ and the second $E_8$ be called 
as $G$. 
Let $\Delta_{G'}( T, U, V)$ and   $\Delta_G(T, U, V)  $ be the 
corresponding one-loop corrections to gauge coupling corrections.  $T, U $ refer to the K\"{a}hler and complex structure moduli of the 
torus $T^2$ and $V$ is the Wilson line modulus in $T^2$. 
Then it was shown  \cite{Stieberger:1998yi} that the difference in the thresholds is given by 
\begin{eqnarray}
\Delta_{G'}( T, U, V) - \Delta_{G} ( T, U, V) =  - 48 \log \left[( {\rm det\, Im } \Omega) ^{10} \left|
\Phi_{10}( T, U, V)  \right|^2  \right]  \ ,
\end{eqnarray}
where 
\begin{equation}
\Omega = \left( \begin{array}{cc}
U & V \\ V & T 
\end{array} \right)  \ ,
\end{equation}
and $\Phi_{10}(T, U, V)$ is the  unique 
Siegel modular form of weight $10$ transforming under the   duality group 
$Sp(2,\mathbb{Z}) \simeq SO(3, 2,  \mathbb{Z}) $. 
In \cite{Stieberger:1998yi}, it was  also shown that this difference in thresholds was independent of the 
way $K3$ was realized and is also holds  for non-standard embeddings. 
In this paper, we evaluate the difference for 
 heterotic compactifications on CHL orbifolds of $K3$ and 
show  that the 
difference in the threshold corrections for  the two gauge groups $G, G'$  is given by 
\begin{eqnarray}\label{diff}
\Delta(G, T, U, V) - \Delta( G', T, U, V) =  - 48\log \left[ ({\rm det\, Im } \Omega) ^{k} \left|
\Phi_{k}( T, U, V)  \right|^2  \right],
\end{eqnarray}
where $\Omega^{k}$ is a weight $k$ modular form 
transforming under subgroups of $Sp(2, \mathbb{Z}) $ with  $k$ 
\begin{equation}\label{defk}
k = \frac{24}{N+1} - 2 \, ,
\end{equation}
where $N=2, 3, 5, 7$ labels the various CHL orbifolds. 
This   generalizes the observation 
in \cite{Stieberger:1998yi}.  Thus the  gauge threshold corrections
are automorphic forms under  sub-groups of the duality group of the   parent  un-orbifolded theory. 

The cusp form $\Phi_{10}$  also makes its appearance in partition function of dyons 
in heterotic on $T^6$, a theory which has ${\cal N}=4$ supersymmetry 
\cite{Dijkgraaf:1996it,LopesCardoso:2004xf,Shih:2005uc,Gaiotto:2005hc} \footnote{It was
recently shown that certain BPS saturated  amplitude in type II on $K3\times T^2$  also depends on 
$\Phi_{10}$ \cite{Hohenegger:2011us}. }. 
This theory is related to type II on $K3 \times T^2$ by string-string duality. 
In \cite{Jatkar:2005bh,justin1,David:2006yn}, it was shown that the partition function of dyons for 
the CHL orbifolds of 
the heterotic preserving ${\cal N}=4$  supersymmetry
 are captured by Siegel modular forms of weight 
$k$  transforming under subgroups of $Sp(2, \mathbb{Z})$  with  $k$ given by (\ref{defk}) 
for the various CHL orbifolds of the heterotic theory.  
These theories are related to  type II on the CHL orbifold of $K3$ which has 
${\cal N}=4$ supersymmetry.
We show that the modular forms $\Phi_k$ obtained for  the difference of the thresholds
in (\ref{diff}) 
are related by a $Sp(2, \mathbb{Z})$ transformation to the dyon partition function 
in CHL orbifolds.  The relationship between the difference in the thresholds 
of the non-abelian gauge groups of the  ${\cal N}=2$ heterotic compactification 
to the dyon partition functions in the ${\cal N}=4$ heterotic  is  certainly interesting and
worth exploring  further.  We will comment on this relation in (\ref{sec:6}).

This paper is organized as follows. In section \ref{sec:2}, we discuss the spectrum of heterotic compactifications
on the CHL orbifold $(K3 \times T^2)/\mathbb{Z}_N$ and show that the orbifold preserves the number of vectors but reduces the 
number of hypers. 
In section \ref{sec:3}, we evaluate the new supersymmetric index for compactifications on the CHL orbifold of $K3$. 
We will discuss the case of $N=2$ in detail for which we realize $K3$ as a $\mathbb{Z}_2$ orbifold. We then generalize 
the results for the other values of $N$. In section \ref{sec:4}, we  show that the 
the new supersymmetric index for these orbifolds   contains representations of  the Mathieu group $M_{24}$. 
In section \ref{sec:5}, we evaluate the difference in the gauge corrections between the groups $G$ and $G'$ and show that 
it is captured by a modular form $\Phi_k$  transforming under subgroups of $Sp(2,\mathbb{Z} )$. 
Section \ref{sec:6} contains our conclusions and discussions. 
Appendix \ref{app:a}  contains various identities involving modular forms used to obtain our results. 
Appendix \ref{app:b} contains details regarding lattice sums and finally appendix \ref{app:c} 
has the details of the calculations for the $\mathbb{Z}_2$ CHL orbifold of $K3$.

 \section{Spectrum of heterotic on CHL orbifolds of $K3$ }\label{sec:2}

 In this section we derive the spectrum on $(K3 \times T^2)/\mathbb{Z}_N$ compactifications. 
 Before we go ahead, let us recall how  these manifolds are constructed. 
 The non-zero hodge numbers of  $K3$ are  given by 
 \begin{equation}
  h_{(0,0)} = h_{(2,2)} = h_{(0,2)}= h_{(2,0)} = 1, \qquad h_{(1,1)} = 20. 
 \end{equation}
The Hodge numbers of $T^2$ are given by 
\begin{equation}
 h'_{00} = h'_{(1, 0)} = h'_{(0, 1)} =  h'_{(1, 1) } = 1. 
\end{equation}
 To ensure ${\cal N}=2$ supersymmetry we need to preserve $SU(2)$ holonomy. 
 This implies that the $\mathbb{Z}_N$  acts freely \cite{Aspinwall:1995fw}. 
 The orbifold action 
 must also preserve the holomorphic 
 $2$-forms on $K3$ and the holomorphic $1$-form   on $T^2$.  
 It is known that  the $\mathbb{Z}_N$ symmetry action on $K3$ always involves fixed points on $K3$ \cite{Nikulin},  
 therefore
 it should freely act  on $T^2$.  This action is just a shift by a unit $1/N$ on one of the circles of 
 $T^2$. Since the orbifold action involves both $K3$ and $T^2$ the compactifications 
 on the CHL orbifold of $K3$
 can not be thought of as obtained from a ${\cal N}=1$ vaccum in $d=6$. 
 Thus $(0,0)$ and $(2, 2)$ form are just the scalar form and the volume form on $K3$ 
which are preserved under the action of $\mathbb{Z}_N$.
Also the the $1/N$ shift on the circle does not project out any of the forms on $T^2$.
Thus the orbifold acts  only on the 
$(1, 1)$-forms of  $K3$.  The number of such  forms on $K3$  which are  invariant 
 are given by  $2k$ with \cite{justin1}
 \begin{equation}
  h_{(1, 1)} = 2k, \qquad k= \frac{24}{N+1} - 2, \qquad {\rm for } \; N= 2, 3, 5, 7. 
 \end{equation}
Among the $(1, 1)$ forms which are not projected out is the K\"{a}hler form $g_{k\bar l}$. 
The K\"{a}hler form, the $(0, 2)$ and $(2,0)$ forms are  self dual while the $2k-1$ forms 
are anti-self dual.  Thus the Euler number of the orbifold along the $K3$ directions 
reduces to  $2k  + 4$. 
 This information of the CHL orbifold  $(K3 \times T^2)/\mathbb{Z}_N$ is sufficient to 
 obtain the spectrum of massless modes in $d=4$. 
 We generalize the method developed in \cite{Walton:1987bu} 
 for $K3$ compactification of the heterotic string. 
We will first discuss the states arising from   compactifying  the  $d=10$ graviton multiplet
 and  then we will examine the spectrum from the $d=10$  Yang-Mills multiplet.

 \subsection*{Universal  sector}
 
 We call the spectrum from the $d=10$ graviton multiplet the universal sector. 
 This multiplet consists of the following fields 
 \begin{equation}
  R(10)  = \{ G_{MN}, \Psi^{(-)}_M, B_{MN}, \Psi^{(+)}, \varphi \} . 
 \end{equation}
 Here  
  $G_{MN}$ is the graviton, $\Psi^{(-1)}$ is a negative-chirality Majorana-Weyl gravitino
 gravitino, $B_{MN}$ the anti-symmetric tensor and $\Psi^{(+)}$ is a positive-chirality 
 Majorana-Weyl spinor. 
 On dimensional reduction these fields should organize themselves to 
 a ${\cal N}=2$ graviton multiplet, vector multiplets and hypermultiplets in $d=4$. 
 The field content of these multiplets are given by 
 \begin{eqnarray}
 R(4) &=& \{ g_{\mu\nu}, \psi^{i}_\mu,  , a_\mu \},  \qquad\quad i = 1, 2 , \\  \nonumber
 V(4) &=& \{ A_\mu, \psi^{\prime i}, \phi^i  \}, \\ \nonumber
 H(4) &=& \{ \chi^i,   \varphi^a  \}, \qquad\qquad a = 1, \cdots 4. 
\end{eqnarray}
The ${\cal N}=2$ graviton multiplet in $d=4$ consists of a graviton $g_{\mu\nu}$, two Majorana
gravitinos  $\psi^i_\mu, \, i = 1, 2, $ and the graviphoton $a_\mu$. 
The vector multiplet consists of the gauge field $A_\mu$, 
two Majorana spinors $\psi^{\prime i}$  and two real  scalars $\phi^i$. 
The hypermultiplet consists of two Majorana spinors $\chi^i$ and $4$ real scalars $\varphi^a$ with 
$a = 1 \cdots 4$.  We will label the $4$ non-compact direction by $\mu, \nu  \in \{ 0, 1, 2, 3\}$. 
The directions of the $T^2$ by $r, s  \in \{ 4, 5 \}$ and the directions of the K3 by 
$m, n \in \{6, 7, 8, 9 \}$.  

Let us first examine  the bosonic fields  under dimensional reduction. The $d=10$ graviton 
reduces as $G_{\mu\nu} = g_{\mu\nu} (x) \otimes 1 \otimes 1 $ where $1$ refers to the constant 
scalar form  on $(K3 \times T^2)/\mathbb{Z}_N$.  
There are $2$ vectors from $ G_{\mu  r} = A_\mu(x)  \otimes f_r \otimes 1$ where 
$f_r$ refers to the $2$ holomorphic  1-forms on $T^2$ which are unprojected 
by the orbifold.  Similarly there are  $2$ vectors $ B_{\mu r} = A_{\mu} (x) \otimes f_r \otimes 1$. 
These $4$ vectors arrange themselves into the  single  graviton multiplet and 
$3$ vector multiplets.  Let us now count the total number of scalars, this will determine the 
number of hypers.  There are totally $4$ scalars from  the following 
components of the  metric in 10 dimensions $G_{44} , G_{55}, G_{45}, B_{45}$.  
Now  consider  the scalars arising from the  metric and the anti-symmetric tensor 
with indices along the $K3$ directions. The anti-symmetric tensor reduces as 
$B_{mn} = \phi(x)\otimes 1\otimes f_{mn}$ where $f_{mn}$ are the   harmonic 
$2$-forms on the CHL orbifold of $K3$.  This results in $2k + 2$ scalars. 
To obtain massless scalars from the metric we require solutions of the Lichnerowicz 
equation on the CHL orbifold of $K3$. These are constructed as follows, 
let us use $a, \bar b \in \{1, 2 \}$ to  refer  to the two complex directions  along 
the CHL orbifold of $K3$. Then  the zero modes  from the metric are constructed 
as follows \cite{Walton:1987bu}
\begin{eqnarray}
h_{a\bar b} &=& f'_{a \bar b}, \\ \nonumber
h_{ab}&=& ( \epsilon_{ac} f'_{b\bar d} + \epsilon_{bc} f'_{a\bar d} ) g^{\bar d c}, \\ \nonumber
h_{\bar a\bar b} &=&  h_{ab}^*. 
\end{eqnarray}
Here $f'_{a \bar b }$ refer to the $2k$ harmonic $(1, 1)$-forms on the CHL orbifold of $K3$. 
Note that $h_{a, b}$ and $h_{\bar a \bar b}$ vanish when  $f'_{a\bar b}$ is the 
K\"{a}hler form. Therefore there are $3 \times 2k - 2$ solutions of the 
Lichnerowicz 
equation on the CHL orbifold of $K3$. This leads to  $6k - 2$ scalars from the dimensional 
reduction of the metric with indices along the CHL orbifold of $K3$. 
The $10$ dimensional  dilaton reduces as $\varphi = \varphi(x) \otimes 1 \otimes 1$ to give 
rise to a single scalar. Finally the  anti-symmetric tensor reduces as 
$B_{\mu\nu} = b_{\mu\nu}(x) \times 1 \times 1$, but a anti-symmetric tensor 
in  $d=4$ is equivalent to a scalar by hodge-duality. 
Adding all the scalars we get $ 8k  + 6 $ scalars. 
Among these $6$ scalars are needed to complete the $3$ vector muliplets. 
The rest of the scalars arrange themselves in to $2k$ hyper multiplets. 
To summarize we have  the following dimensional reduction 
of the graviton multiplet in  $d=10$. 
\begin{equation}\label{univ}
R(10) \rightarrow R(4) + 3 V(4) + 2k H(4) \, . 
\end{equation}

To complete the analysis let us verify that the fermions also arrange themselves 
into these multiplets. 
Before we go ahead we need to recall some facts about index theory. 
There is a one to one correspondence of solution of the massless Dirac equation 
on a $4$ dimensional  complex manifold and the number of harmonic $(0, p)$ forms \cite{Candelas:1985en,Witten:1985xc}.  
The $(0, 0)$ form and a $(0, 2)$ form on the CHL orbifold of $K3$ 
results in two real Dirac zero modes  which have negative internal chirality \cite{Walton:1987bu}. 
Let us call these spinors $\Omega$ and $\omega$. 
Consider the gravitino in $d=10$ it reduces to a Rarita-Schwinger field 
in $d=4$ as the following $4$ real  gravitinos
\begin{eqnarray}\label{gravred}
\Psi_\mu^{(-)}  &=& \psi_\mu^{(+)1}  (x) \otimes \xi^{(+)} \otimes    \Omega^{(-)}, \\ \nonumber
\Psi_\mu^{(-)}  &=& \psi_\mu^{(-)1}  (x) \otimes \xi^{(-)} \otimes    \Omega^{(-)}, \\ \nonumber
\Psi_\mu^{(-)}  &=& \psi_\mu^{(+)2}  (x) \otimes \xi^{(+)} \otimes    \omega^{(-)}, \\ \nonumber
 \Psi_\mu^{(-)}  &=& \psi_\mu^{(-)2}  (x) \otimes \xi^{(+)} \otimes    \omega^{(-)}, 
 \end{eqnarray}
 where $\xi^{(\pm) }$ are the constant spinors on $T^2$.  The superscripts refer to the 
 chirality.  These $4$ real spinors organize themselves as $2$   Majorana 
Rarita-Schwinger fields $\psi_\mu^{i}$  in $d=4$.
These form the superpartners in the graviton multiplet 
$R(4)$. 
Now consider again  the gravitino in $10$ dimensions  and reduce it with the 
vector index along the $T^2$ directions, these result in spinors in $d=4$.   Using the similar reduction as in 
(\ref{gravred}) we can conclude that there are $2  \times 2 = 4$  Majorana spinors  in $d=4$. 
Finally reduce the  $d=10$ spinor $\Psi ^{(+)}$  again on similar lines as in (\ref{gravred}) and 
we obtain $2$ Majorana spinors in $d=4$.  Thus totally we have $6$ Majorana spinors 
which form the superpartners of the $3$ vectors multiplets. 
Now let us move to the situation when the gravitino has indices along the CHL orbifold of $K3$. 
Now given a harmonic $(1, 1)$ form we can construct the following solutions to the 
Rarita-Schwinger equations  on the CHL orbifold of $K3$ \cite{Walton:1987bu}. 
\begin{eqnarray}
\zeta_a = f'_{a\bar b}\Gamma^{\bar b} \Omega^{(-)} , 
\qquad 
\zeta_{\bar b} = f'_{a\bar b} \Gamma^a \omega^{(-)}. 
\end{eqnarray}
 Here $\Gamma$'s are the internal $\gamma$-matrices and $f'$ refer to the $2k$ 
 $(1,1)$ forms. Again by reducing the $d=10$ gravitinos   with a similar construction 
 as in (\ref{gravred}) but with the vector indices of the gravitino  along the  CHL orbifold of
 $K3$ we obtain $2 \times 2k = 4k$ Majorana spinors in $d=4$
   which  form the fermionic content  in the 
 $2k$ hyper multiplets. 
 This completes the analysis of the dimensional reduction of the graviton 
 multiplet in $10$ dimensions  which results in the fields given in (\ref{univ}). 
 Thus we see that  it is only the number of hypers in the universal sector 
 which is sensitive to the orbifolding.

\subsection*{Gauge sector}

Now let us examine the spectrum that arise from dimensional reduction of the Yang-Mills multiplet in 
$d=10$. The field content of this multiplet  is given by 
\begin{equation}
 Y(10) = \{ A_M, \Lambda^{(-)} \}. 
\end{equation}
The negative chirality Majorana fermions as well as the gauge bosons
are  in the adjoint representation of $E_8\otimes E_8$ transforming as 
$({\bf 248},  {\bf 1} ) \oplus ( {\bf 1} , {\bf 248 } )$.  This multiplet must decompose to  ${\cal N}=2$ 
vectors and hypers in $d=4$. To obtain the number of vectors and hypers we will use 
index theory to find the number of zero modes of fermions in the CHL orbifold of $K3$. 
To preserve supersymmetry    in $d=4$ the spin connection must be  set to equal to the gauge connection. 
 Let us consider the standard embedding in which the we take an $SU(2)$ out of the first $E_8$  and set it equal to the
 spin connection on the CHL orbifold of $K3$. As mentioned earlier the $SU(2)$ holonomy of the spin connection 
 is preserved by the orbifolding procedure. 
 This procedure breaks the $E_8$ to a subgroup, let us consider   the maximal subgroup $E_7 \otimes SU(2) $,  
in which the  $SU(2)$  of the gauge connection is set equal to the $SU(2)$ spin connection. 
Under the maximal subgroup $E_7 \otimes SU(2)\otimes E_8 $, the Yang-Mills multiplet decomposes as follows. 
\begin{eqnarray} \label{decomp}
 ({\bf 248} , {\bf 1} ) \oplus ( {\bf 1} , {\bf 248} ) = ( {\bf 133} , {\bf 1} , {\bf 1} ) 
 \oplus ( {\bf 1} ,{\bf  3} , {\bf 1} ) \oplus ( {\bf 56} , {\bf 2} , {\bf 1} ) \oplus ({\bf  1} , {\bf 1} , {\bf 248} ) . 
\end{eqnarray}
On the left hand side of the above equation we have kept track of the quantum numbers of $E_7, SU(2)$ and the second 
$E_8$. 
Dimensional reduction  of the $d=10$  gauge bosons in the  
$( {\bf 133} , {\bf 1} , {\bf 1} )\oplus ({\bf  1} , {\bf 1} , {\bf 248} )$
representation 
to $d=4$ gives rise to gauge bosons in 
the $({\bf 133}, {\bf 1}) \oplus ( {\bf 1}, {\bf 248}) $  representation of $E_7\otimes E_8$. 
The corresponding scalars in these vector multiplets also arise in the dimensional 
reduction from  the $d=10$ gauge bosons with vector indices along the $T^2$ directions.  
Now the fermionic super partners of these fields in the vector multiplets arise as follows. 
Consider the fermions of  Yang-Mills multiplet in $d=10$ in the representation 
$ ( {\bf 133}, {\bf 1} , {\bf 1} ) \oplus ( {\bf 1}, {\bf 1}, {\bf 248} )  $ , they are uncharged 
respect to the $SU(2)$ and therefore  behave conventionally. 
That is for these fermions, we can use the two spin $1/2$ zero modes on the CHL orbifold of $K3$  
of negative chirality denoted by 
$\Omega, \omega$  earlier to 
to construct   two Majorana fermions in $d=4$ in the same representations. 
These  are the fermionic partners in the vector multiplets. 
Let us state the existence of the two spin $1/2$ zeros modes as an index theorem. 
Essentially we have 
\begin{eqnarray} \label{euler}
 I_{\gamma\cdot\nabla} = n_{1/2}^{(-1) } - n_{1/2}^{(+1)} = \frac{1}{ ( 2k + 4 ) ( 8\pi^2) } \int {\rm Tr} ( R\wedge R) = 2. 
\end{eqnarray}
Note that, we have normalized the integral by the Euler number of the CHL orbifold and the integral is also performed over the 
orbifold. 
$n_{1/2}^{(\pm 1)}$ counts the number of massless spin $1/2$ zero modes of the 
appropriate chirality.

Let us examine the fermions which are charged under the  $SU(2)$
in the decomposition (\ref{decomp}).
Since the corresponding gauge connection is  identified to be the spin connection, these fermions
must arrange themselves into ${\cal N}=2$ hypers. 
First consider the    fermions  which transform  non-trivially under the  $SU(2)$. 
To obtain the number of fermions in $d=4$ we need to use the index theorem
of the Dirac operator on the of the CHL orbifold of $K3$.  
Since these fermions are charged under the $SU(2)$ we need the expression for the twisted index, which 
is given by  \cite{Eguchi:1980jx}
\begin{eqnarray}\label{twind} 
 I_{\gamma \cdot \nabla}^{{\bf r} }   &=&  n_{1/2}^{(-1) }({\bf r} ) - n_{1/2}^{(+1)} ({\bf r} ) , \\ \nonumber
 & = & \frac{1}{8\pi^2} \int \left(  \frac{{\bf r} }{( 2k + 4) } {\rm Tr} ( R\wedge R) - {\rm Tr} _{{\bf r}}  ( F\wedge F)  \right) .
\end{eqnarray}
Here ${\bf r}$ is the representation of the fermions. 
Note that just as in (\ref{euler}) we have normalized the integral of the curvature term by the Euler number of the 
CHL orbifold of $K3$.  For  $k=10$, the expression reduces to that for $K3$. 
Setting the gauge connection equal to the spin connection we obtain 
\begin{equation}
 {\rm Tr}_{\bf 2}  ( F \wedge F) = \frac{1}{2} {\rm Tr} ( R\wedge R) \, . 
\end{equation}
The $1/2$ is because the trace in the ${\rm Tr} ( R\wedge R) $ is taken in the ${\bf 4}$ of $SU(4)$ which are two doublets of 
$SU(2)$. 
Now  one can relate the trace in representation ${\bf r}$ to the trace in the doublet by 
\begin{equation}
 {\rm Tr}_{\bf r}  ( F\wedge F) =\frac{1}{6} {\bf r}  ( {\bf r} ^2 - 1)  {\rm Tr}_{{\bf 2}}   ( F \wedge F) \, . 
\end{equation}
Substituting this relation in (\ref{twind}) and using the last equality in (\ref{euler}) we obtain 
\begin{equation} \label{twind2}
  n_{1/2}^{(-1) }({\bf r} ) - n_{1/2}^{(+1)}({\bf r} )  = 2{\bf r}  - \frac{1}{3} ( k+2) {\bf r}  ( {\bf r} ^2 -1)  \, .
\end{equation}
Note that for the singlet ${\bf r} =1$, the expression shows that there exist two negative chirality modes which 
was known by explicit construction as the spinors $\Omega^{(-1)}, \omega^{(-1)}$. 
Now each pair of spin $1/2$ zero modes  given by the index  (\ref{twind2}) gives rise to 
a pair of Majorana fermions in $d=4$ which form the fermions in a single hypermultiplet. 
Thus the number of hypers  in the representation ${\bf r}$ of $SU(2)$  in $d=4$ from the gauge 
sector is given by 
\begin{equation} \label{nuhyp}
 N_H^{{\bf r}} = \frac{1}{6} (k+2) {\bf r}  ( {\bf r} ^2 -1) - {\bf r }  \, .
\end{equation}
Note that this is always an integer. 
Let us apply this formula to the  fermions which transform non-trivially under $SU(2)$. 
Consider the doublets transforming as $( {\bf 56} , {\bf 2} , {\bf 1} )$. 
Using (\ref{nuhyp})  we can conclude that there are $k$  charged hypers in the 
$( {\bf 56}, {\bf 1} ) $ representation of $E_7\times E_8$. 
Similarly consider the triplets $( {\bf 1} ,{\bf  3} , {\bf 1} )$ which lead to  
$4(k+2) -3$ hypers uncharged under the gauge group. 
From the above discussion we see that the Yang-Mills multiplet in $d=10$ results in the following 
multiplets in $d=4$
\begin{eqnarray}
 Y(10) \rightarrow & &   V(4) [ ({\bf 133} , {\bf 1} ) + ( {\bf 1} , {\bf 248} ) ]    \\ \nonumber
 & & +
 H(4) [  k ({\bf 56} , {\bf 1} ) +  ( 4 ( k +2) - 3) ( {\bf 1} , {\bf 1} )  ] \, .
\end{eqnarray}
Here we have also indicated  the representations of $E_7\otimes E_8$.
As a simple check note that for $K3$ we have $k=10$ which results in  the well known  10 charged hypers and 
65 uncharged hypers \cite{Kachru:1995wm}. 
The complete spectrum in $d=4$ is given by 
\begin{eqnarray}
R(10) + Y(10) \rightarrow  & & R(4) +   V(4) [ 3 ({\bf 1} , {\bf 1} )  + ({\bf 133} , {\bf 1} ) + ( {\bf 1} , {\bf 128} ) ]  \\ \nonumber
& &+  H(4) [ k ( {\bf 56} , {\bf 1} )  + (  6k  + 5  ) ( {\bf 1} , {\bf 1} ) ] \, .
\end{eqnarray}
Thus, compactifications on the CHL orbifold of $K3$ change the number of the hypers. 
It is important to note that these orbifolds involve the shift on $S^1$ together with the 
involution in $K3$ which reduces the number of $(1, 1)$ forms. Therefore, they cannot be 
thought of as  a four manifold which implies this compactification cannot be lifted 
to $6$ dimensions.  Thus, the difference in the number of hypers and vectors is not constrained 
by anomaly cancellation in $d=6$. 

Let us now discuss the generic spectrum of these models.
The generic spectrum is labeled  by the number of uncharged hypers $M$  and number of commuting 
$U(1)$  denoted by $N$. For the embedding of $SU(2)$ we have considered the model is 
given by 
\begin{equation}
(M, N) = ( 6k + 5 , 19). 
\end{equation}
We have listed this for the various $(M, N)$  values of $k$ corresponding to the CHL orbifold. 
\begin{eqnarray}
k= 10, & &    \qquad \qquad ( 65, 19 ) ,  \\ \nonumber
k =  6, & & \qquad \qquad ( 41, 19 ) , \\ \nonumber
k =  4, & & \qquad\qquad ( 29, 19 ) ,  \\ \nonumber
k =  2, & &  \qquad \qquad ( 17, 19 ) , \\ \nonumber
k = 1, & &  \qquad \qquad (11, 19 ).  
\end{eqnarray}
For all of these models the unbroken gauge group is $E_7 \otimes E_8$. 
In the dual type II theory these models arise from Calabi-Yau compactifications with 
Hodge numbers $(h_{(1,1)}, h_{(2, 1)} ) = (N-1, M-1) = (18, 6k + 4)$. 
CHL orbifolding of $K3$ just reduces the number of hypers. 

Let us now consider compactifications in which a  $SU(n)$ with $n= 3, 4, 5$  of one of the $E_8$ is 
embedded in the spin connection. 
Doing so, breaks the  $E_8$ to $E_6, \ SO(10)$ and  $SU(5)$ respectively. 
The number of uncharged hypers from the gravition multiplet remains invariant 
and is given by $2k$.
A similar analysis shows that 
the number of uncharged hypers from the Yang-Mills multiplet is given by the 
index
\begin{equation}
N_H({\rm singlets})  = (2k + 4) n - ( n^2 -1) . 
\end{equation}
Note that this expression reduces to $ 4 ( k+ 2) - 3$ for $n=2$ as seen earlier
in detail. Therefore  adding the $2k$ uncharged hypers from the 
universal sector,  the total  number of uncharged hypers for these compactifications 
is given by  $ 2k (n+1)  - ( n^2 - 4n - 1) $. 
Thus the $(M, N) $ values for these models are
\begin{equation}
 (M, N) = ( 2k [n+1]  - [ n^2 - 4n - 1] , 21 - n  ). 
\end{equation}
Again we see that it  is only the  number of hypers that are affected by $k$.
These models are the generalization of the ones considered in \cite{Kachru:1995wm} for $k=10$. 
Though the number of vectors are not affected by these compactifications, it will be clear from 
our analysis of the threshold corrections that the duality group under which these models are invariant
are subgroups of the parent theory.

\def\CHL{{\rm CHL}}

\section{New supersymmetric index for CHL orbifolds of $K3$}\label{sec:3}

In this section we evaluate the new supersymmetric index for the CHL orbifold of $K3$. 
This index forms the basic ingredient for both gauge and gravitational threshold corrections 
for the heterotic compactifications we considered in the previous section. 
The new supersymmetric index is defined as \footnote{We will use $q, \tau$ to refer to the modular parameter
of the worldsheet, they are related by $q= e^{2\pi i \tau}$ and similarly $\bar q= e^{-2\pi i \bar \tau}$. }
 \begin{equation} \label{neind1}
 {\cal Z}_{\rm new} 
  (q, \bar q )  = \frac{1}{\eta^2(\tau)  } {\rm Tr}_{R} 
 \left(   F e^{i \pi  F} q^{L_0 - \frac{c}{24} } \bar q^{\bar L_0 -\frac{ \bar c}{24}}   \right) . 
\end{equation}
Here, the trace is taken over the internal CFT with central charge $( c, \tilde c ) = ( 22, 9 )$. 
Note that the left movers are bosonic while the right movers are supersymmetric. The right moving
internal CFT has a ${\cal N}= 2$ superconformal symmetry.  It admits a $U(1)$ 
 current which can serve as the world sheet fermion number, we  denote this
as $F$. The subscript $R$ refers to the fact that we take the trace in the Ramond sector for the 
right movers.
For the $K3\times T^2$ compactifications, this index was evaluated in \cite{Harvey-Moore} using the
$\mathbb{Z}_2$ orbifold realization of $K3$. 
We will first generalize this  computation for the CHL orbifold $(K3 \times T^2)/\mathbb{Z}_2$. 
Then using  observations from  the explicit calculations done for the $\mathbb{Z}_2$ orbifold,
we will generalize and obtain the expression  of  the new supersymmetric index 
for the 
CHL orbifolds $(K3 \times T^2)/\mathbb{Z}_N$ with $N= 3, 5, 7$. 

\subsection{The $\mathbb{Z}_2$ orbifold}

The $N=2$ CHL orbifold of $K3$ 
admits the following  simple orbifold realization. 
First, $K3$ is realized as a $\mathbb{Z}_2$ orbifold by the action $g$  on a torus $T^4$, and  then, the 
CHL orbifold of $K3$ is obtained by the action of $g'$ given below. 
\begin{eqnarray}\label{orbiact}
 g:\quad & ( y^4, y^5, y^6, y^7, y^8, y^9) \rightarrow   ( y^4, y^5, - y^6, -y^7, -y^8, -y^9) , \\ \nonumber
 g':\quad & ( y^4, y^5, y^6, y^7, y^8, y^9 ) \rightarrow ( y^4 +\pi, y^5, y^6+ \pi , y^7, y^8, y^9)  .
\end{eqnarray}
Here, the directions $4, 5$ label the $T^2$ and the $6, 7, 8, 9$ directions are the 
$K3$ directions. Note that, the $g'$ action involves as shift of $\pi$ along one of the circle of $T^2$. 
This is embedded in the heterotic string by  performing a shift of $\pi$ along $2$ of the directions of the 
$E_8'$ lattice \footnote{The lattice 
in which the spin connection is embedded will be denoted by $E_8^\prime$ or 
$G'$. } i.e.~there is a shift  given by 
\begin{equation}\label{lshift}
 X^I \rightarrow X^I + ( \pi, \pi, 0, 0 ,0 , 0 , 0, 0) ,
\end{equation}
where $X^I$ refer to the bosonic co-ordinates of the $E_8^\prime $ lattice. 
If the action $g'$ is not implemented the action of $g$ together with the 
shift in (\ref{lshift}) breaks $E_8'$ to $E_7$. The presence of $g'$ ensures the 
CHL orbifolding. 
This shift  in (\ref{lshift}) is coupled to the $g, g'$ action as follows. 
\begin{eqnarray} \label{newindex1}
 {\cal Z}_{\rm new} 
  (q, \bar q )  & =&   \left( 
  \frac{1}{\eta^2(\tau) } \sum_{a, b = 0, 1} 
  {\cal Z}_{( a, b) } [ E_8'; q] \times  {\cal Z}_{(a, b) } [ \text{CHL} ;  q, \bar q ]  \right)\times  {\cal Z} [E_8 ; q ] .
\end{eqnarray}
Here, ${\cal Z} [ E_8;  q ] $ is the partition function of the  second $E_8$ lattice which is given by 
\begin{equation}
 {\cal Z} [ E_8 ;  q ]   = \frac{E_4}{\eta^8 } .
\end{equation}
The Eisenstein series, $E_4$, admits the following decomposition in terms of theta functions.  
\begin{equation}
 E_4 = \frac{1}{2} \left( \theta_2^8 + \theta_3^8 + \theta_4^8 \right) . 
\end{equation}
The partition function of the $E_8'$ which involves the following  shifted lattice sum. 
\begin{equation}
 {\cal Z}_{( a, b) } [ E_8'; q] = 2 \frac{1}{\eta^8}  e^{- 2\pi i\frac{ ab}{n^2} \gamma^2 }
 \sum_{\lambda  \in \Gamma^{8}+ \frac{a}{2}\gamma}
e^{2\pi i \frac{b}{n}\lambda \cdot \gamma} q^{\frac{1}{2}\lambda^2}.
\end{equation}
The sum runs over all the lattice vectors $\lambda$ of $E_8$. 
The lattice shift $\gamma $  for the $\mathbb{Z}_2$  case   is given by  
\begin{equation}
 \gamma  = ( 1, 1, 0, 0, 0, 0, 0, 0), \qquad n = 2 \ .
\end{equation}
In appendix  \ref{app:b} we have evaluated the shifted lattice sum for various  
 values of $(a, b)$. This result is given by  
 \begin{align} \label{shifte8}
{\cal Z}_{(0,0)} [E_8'; q ] &= \frac{\th_2^8 + \th_3^8 + \th_4^8}{\eta^{8}}, \qquad 
{\cal Z}_{(0,1)}[E_8'; q] = \frac{\th_3^6 \th_4^2 + \th_4^6 \th_3^2}{\eta^{8}} ,\\  \nonumber
{\cal Z}_{(1, 0 )}[ E_8'; q] &= \frac{\th_2^6 \th_3^2 + \th_3^6 \th_2^2}{\eta^{8}}, \qquad \  
{\cal Z}_{(0,1)} [E_8'; q] = - \, \frac{\th_2^6 \th_4^2 - \th_4^6 \th_2^2}{\eta^{8}} .
\end{align}
 What is now left, is to define the partition function over $( K3\times T^2)/ \mathbb{Z}_2$
 referred as $ {\cal Z} [ \CHL; q, \bar q]  $ in (\ref{newindex1}).  
 For this we first define the lattice momenta on the $T^2$ which is given by 
 \begin{align}\label{plpr}
  \frac{1}{2}{p_R^2}  &= \frac{1}{2  T_2 U_2} | - m_1 U +m_2 + n_1 T + n_2 TU |^2 , \\
  \nonumber
  \frac{1}{2}{p_L^2} &= \frac{1}{2} p_R^2  + m_1 n_1 + m_2 n_2  \, . 
 \end{align}
 The variables $T, U$ refer to the complex structure and the K\"{a}hler moduli of the torus $T^2$.  
Then the partition function can be written as 
\begin{equation}\label{indchl}
 {\cal Z}_{(a, b) } [ \CHL; q, \bar q] = \frac{1}{\eta^2} \sum_{m_1, m_2, n_1, n_2} 
 q^{\frac{1}{2} p_L^2} \bar q^{ \frac{1}{2} { p_R^2} }  
 {\cal F} _{m_1, m_2, n_1, n_2} (a, b ; q), 
\end{equation}
where the $1/\eta^2$ factor arises due to the left moving bosonic oscillators   
where ${\cal F}_{m_1, m_2, n_1, n_2} (a, b ; q) $ is independent of $T, U$ and is given by 
\begin{eqnarray} \label{tracdef}
 {\cal F}_{m_1, m_2, n_1, n_2} (a, b; q)   &=& \frac{1}{2}  \sum_{r, s  = 0}^{1} 
 F_{m_1, m_2, n_1, n_2} ( a, r, b,  s; q ) , \\ \nonumber 
 F_{m_1, m_2, n_1, n_2} ( a, r, b, s ; q) & =& 
 {\rm Tr}_{m_1, m_2, n_1, n_2; g^a , g^{\prime r};  RR} \left(  g^b g^{\prime s}
 e ^{ i\pi (F^{ T^4} +  F^{ T^2 })  } 
  (F^{T^4 } +  F^{ T^2 })  q^{L_0'} \bar q ^{\bar L_0'} \right). 
\end{eqnarray}
Here 
\begin{equation}
 L_0' = L_0 - \frac{p_L^2}{2}, \qquad\qquad  \bar L_0' = L_0 - \frac{p_R^2}{2}. 
\end{equation}
The trace refers the trace over subspace of Hilbert space carrying momentum 
$(m_1, m_2)$ and winding $(n_1, n_2) $. The subscripts  $g, g^{\prime}$ in the trace indicates that the trace 
should be taken in the twisted section. 
 The definition of 
$L_0', \bar L_0'$ ensures that the  partition function $ {\cal F} _{m_1, m_2, n_1, n_2}$ is 
independent of the $T^2$ moduli.   Since the left moving bosonic oscillators 
on $T^2$ has been taken into account in (\ref{indchl}), the trace does not involve these oscillators. 
Note that if one does not have the presence  the insertions of the action of the  $\mathbb{Z}_2$  element   $g'$ 
which is responsible for orbifolding  $K3 \times T^2$ , 
 the coupling of the shifts in the  $E_8'$ reduces to the
coupling of $K3$ realized as a involution of $T^4$ by the action of $g$. 
$F^{T^4}$ is right moving  world sheet fermion number 
 of the $(0,4)$ superconformal algebra of $T^4$. This $U(1)$
is twice the $U(1)$ of the $SU(2)$  present in the $(0, 4)$ superconformal algebra. Finally  
$F^{T^2}$ is the right  moving  world sheet  fermion number  of the $(0, 2)$ superconformal algebra of $T^2$.  
It can be seen that among that unless the fermionic zero modes on $T^2$ are saturated the trace 
given in the last line of (\ref{tracdef}) vanishes. Therefore we obtain 
\begin{equation}
 F_{m_1, m_2, n_1, n_2} ( a, r, b,s ; q)  = 
 {\rm Tr}_{m_1, m_2, n_1, n_2; g^a , g^{\prime r};  RR} \left(  g^b g^{\prime s}
 e ^{ i\pi (F^{ T^4} +  F^{ T^2 })  } 
   F^{ T^2 }  q^{L_0'} \bar q ^{\bar L_0'} \right). 
\end{equation}
The detailed evaluation of the trace is provided in the appendix  \ref{app:c}. 
The result for the various sectors are given by 
\begin{eqnarray}\label{chlotr}
  {\cal F}_{m_1, m_2, n_1, n_2}( 0, 0; q)      &=&  0 , \\ \nonumber
  {\cal F}_{m_1, m_2, n_1, n_2} ( 0, 1;q)  & = & \begin{cases}
  - 2 \left(1+(-1)^{m_1}\right ) \frac{\th_3 ^2 \th_4 ^2}{\eta^{4}} \quad   &\text{for } \{m_1, m_2, n_1, n_2\}  \in  \mathbb{Z}, \\
    0  \quad  & \text{for }  \{m_1, m_2, n_2\}  \in  \mathbb{Z}, \;\;  \{n_1\} \in \mathbb{Z } + \frac{1}{2},
 \end{cases}\\ \nonumber
 {\cal F}_{m_1, m_2, n_1, n_2} ( 1, 0 ;q) & =&  \begin{cases} 
 2  \frac{\th_2^2 \th_3^2}{\eta^4}   \quad   &\text{for } \{m_1, m_2, n_1, n_2\}  \in  \mathbb{Z},  \\
  2  \frac{\th_2^2 \th_3^2}{\eta^4}  \quad  & \text{for }  \{m_1, m_2, n_2\}  \in  \mathbb{Z},  \{n_1\} \in \mathbb{Z } + \frac{1}{2}, 
\end{cases}
\\ \nonumber
{\cal F}_{m_1, m_2, n_1, n_2} ( 1, 1 ;q) &=& 
 \begin{cases}
-  2  \frac{\th_2 ^2 \th_4 ^2}{\eta^{4}} \quad   &\text{for } \{m_1, m_2, n_1, n_2\}  \in  \mathbb{Z},  \\
 - 2  (-1)^{m_1} \frac{\th_2 ^2 \th_4 ^2}{\eta^{4}}  \quad   &\text{for }
 \{m_1, m_2, n_2\}  \in  \mathbb{Z}, \;\;  \{n_1\} \in \mathbb{Z } + \frac{1}{2}. 
 \end{cases} 
\end{eqnarray}
The contributions in which the winding $n_1$ takes half integer values arise due to the twisted sectors in the element $g'$. 
The contributions proportional to $(-1)^{m_1}$ arise due to the insertions of the element $g'$ in the trace. 
Note that if one ignores the contributions where $n_1$ takes half integer values  and 
the ones proportional to $(-1)^{m_1}$,  the result for the various sectors is proportional to that for  $K3$  realized 
as a $\mathbb{Z}_2$ orbifold of $T^2$. 
The expressions in (\ref{chlotr})  can be then be substituted in (\ref{indchl}) to obtain the
partition function on the CHL orbifold of $K3$. 

Let us now use the results in (\ref{shifte8}) and (\ref{chlotr}) to obtain the  new supersymmetric index 
given in (\ref{newindex1}). 
Note that the dependence of the traces in (\ref{chlotr})  over the winding and momenta 
is mild. One just needs to  consider the case when $n_1 \in \mathbb{Z }$ and $n_1\in \mathbb{Z } + \frac{1}{2}$ separately.  
Multiplying the various sectors and summing over the sectors we obtain 
\begin{eqnarray}\label{fullans}
 {\cal Z}_{ {\rm new} }^{(2)}  ( q , \bar q ) = \frac{2 E_4}{\eta^{12} }\times \left[  \sum_{m_1, m_2, n_1 n_2 \in \mathbb{Z }}
 q^{\frac{p_L^2}{2} } \bar q ^{\frac{p_R^2}{2}}  
 \left( - 2 \frac{E_6}{ \eta^{12}}  - ( -1)^m \frac{\th_4^4 \th_3^4 (\th_4^4+\th_3^4)}{\eta^{12}} \right) \right.   \\ 
 \nonumber
   \left.  \qquad 
 +\sum_{m_1, m_2,  n_2 \in \mathbb{Z }, n_1 \in \mathbb{Z} + \frac{1}{2} } q^{\frac{p_L^2}{2} } \bar q ^{\frac{p_R^2}{2}}
 \left( 
 \frac{\th_2^4}{\eta^{12}} \left\{ \th_3^4 (\th_2^4+\th_3^4)	+ (-1)^m \th_4^4(\th_2^4-\th_4^4)	\right\} \right) 
 \right]. 
\end{eqnarray}
The superscript ${}^(2)$ 
refers to the fact that this is the index for the orbifold $( K3\times T^2)/\mathbb{Z}_2$. 
Here we have used the decomposition of $E_6$ in terms of $\theta$-functions which is given by 
\begin{equation}
 2E_6 = - \theta_2^6 ( \theta_3^4 + \theta_4^4) \theta_2^2 + \theta_3^6 ( \theta_4^4 - \theta_2^4) \theta_3^2 
 + \theta_4^6 ( \theta_2^4 + \theta_3^4) \theta_4^2 . 
\end{equation}
Note that this is the generalization of the new supersymmetric index obtained for the standard embedding 
in $K3 \times T^2$ compactifications given in (\ref{fact}) 
for which we obtain the  just the term involving $E_6$ in  the  first 
line  (\ref{fullans}).   The result we have in (\ref{fullans}) is  the expression for the 
new supersymmetric index for the compactifications on $(K3\times T^2) /\mathbb{Z}_2$. 

We will now discuss two equivalent ways of rewriting the expression in (\ref{fullans}) which are 
useful for the 
questions addressed in this paper. 

\subsubsection* {Decomposition in terms of characters of $D_6$}

From the general arguments in \cite{Harvey-Moore}, we expect that the new supersymmetric index 
for $K3\times T^2$ 
decomposes in terms of characters of the sub-lattice $D_6$  of $E_8'$. 
The coefficients in this decomposition can be written in terms of the elliptic genus of the 
${\cal N}=4$ superconformal field theory of the  $d=4$ compact manifold. 
For  $K3\times T^2$ compactifications, this  decomposition of the new supersymmetric 
index is given in (\ref{fact}). 
We will show that the new supersymmetric index for the $(K3\times T^2) /\mathbb{Z}_2$
also can be decomposed in terms of characters of $D_6$ with coefficients 
as the 
twisted elliptic genus of $K3$. 
Let us first define the twisted elliptic genus for the CHL orbifolds of $K3$. Let $g'$ be the 
generator of the $\mathbb{Z}_N$ action on $K3$  which results in the CHL orbifold. 
We define the twisted elliptic genus of $K3$ as 
\begin{eqnarray}
F^{(r,s)}(\tau, z) &=& \frac{1}{N} {\rm Tr}^{K3}_{RR; g^{\prime r} } 
\left( ( -1) ^{F^{K3} + \bar F^{K3} } g^{\prime s} e^{2\pi i z F^{K3} } 
q ^{L_0 -c/24} \bar q ^{\bar L_0 - c/24} 
\right),  \nonumber \\ & & \qquad\qquad\qquad\qquad\qquad\qquad  0 \leq r, s, \leq (N-1). 
\end{eqnarray}
where the trace  is taken  in the ${\cal N}=4$ super conformal field theory 
associated with $K3$  in the  $g^{\prime r }$ twisted Ramond sector.  $F^{K3}$ and 
$\bar F^{K3}$ denote the left and right world sheet fermion number
which can be written as the $U(1)$  charges corresponding to the 
$SU(2)$ R-symmetry   in this theory. 
The twisted elliptic genus for the various CHL orbifolds were provided in \cite{justin1}. 
The results for the $N=2$ CHL orbifold are given by 
\begin{eqnarray}
F^{(0, 0) }(\tau, z) & =&  4 \left[ \frac{\theta_2( \tau, z) ^2}{\theta_ 2( \tau, 0 )^2 } + 
\frac{\theta_3( \tau, z) ^2}{\theta_ 3( \tau, 0 )^2 } 
+ \frac{\theta_4( \tau, z) ^2}{\theta_ 4( \tau, 0 )^2 } \right] , \\ \nonumber
F^{(0, 1) } ( \tau, z) &=& 4 \frac{\theta_2( \tau, z) ^2}{\theta_ 2( \tau, 0 )^2} , \; \  \
F^{(1, 0) } ( \tau, z) = 4 \frac{\theta_4( \tau, z) ^2}{\theta_ 4( \tau, 0 )^2},  \; \ \
F^{(1, 0 ) } ( \tau, z) = 4 \frac{\theta_3( \tau, z) ^2}{\theta_ 3( \tau, 0 )^2} . 
\end{eqnarray}
Using these expressions for the twisted elliptic genus we can see that the 
new supersymmetric index in (\ref{fullans})  can be written as   
\begin{align} \nonumber   
 {\cal Z}_{new} ( q , \bar q )^{(2)} &= \frac{2 E_4}{\eta^{12} }\times \left[  \sum_{m_1, m_2, n_1 n_2 \in \mathbb{Z }}
 q^{\frac{p_L^2}{2} } \bar q ^{\frac{p_R^2}{2}}  
 \left\{
  \frac{\theta_2^6}{\eta^6}  \left( 
 F^{(0, 0) }( \tau, \tfrac{1}{2} )  + (-)^m F^{(0, 1)} ( \tau, \tfrac{1}{2} ) \right) \right.  \right.  \\ \nonumber
 & \qquad \qquad\qquad\qquad \qquad    + q^{1/4}   \frac{\theta_3^6}{\eta^6}   
 \left( 
 F^{(0, 0) }( \tau, \tfrac{1+ \tau}{2} )  + (-)^m F^{(0, 1)} ( \tau, \tfrac{1+ \tau}{2} ) \right)
  \\ \nonumber
& \qquad\qquad\qquad  \qquad\qquad\left.  - q^{1/4}   \frac{\theta_4^6}{\eta^6}   
 \left( 
 F^{(0, 0) }( \tau, \tfrac{ \tau}{2} )  + (-)^m F^{(0, 1)} ( \tau, \tfrac{ \tau}{2} ) \right)
 \right\}    \\ \nonumber 
 &\qquad \left.   \hspace{-3cm}+ \hspace{-.7cm} \sum_{ {\substack{ {m_1, m_2, n_2 \in  \, \mathbb{Z }},   {n_1 \in \, \mathbb{Z} + 1/2} } } }
\hspace{-.7cm} q^{\frac{p_L^2}{2} } \bar q ^{\frac{p_R^2}{2}}  
 \left\{
  \frac{\theta_2^6}{\eta^6}   \left( 
 F^{(1, 0) }( \tau, \tfrac{1}{2} )  + (-)^m F^{(1, 1)} ( \tau, \tfrac{1}{2} ) \right) \right. \right.   \nonumber \\ 
 & \qquad\qquad\qquad \qquad 
 + q^{1/4}  \frac{\theta_3^6}{\eta^6}  
 \left( 
 F^{(1, 0) }( \tau, \tfrac{1+ \tau}{2} )  + (-)^m F^{(1, 1)} ( \tau, \tfrac{1+ \tau}{2} ) \right)\nn  \\  
 \label{fullans1} 
  &  \qquad\qquad\qquad\qquad  \left. \left. 
 - q^{1/4}  \frac{\theta_4^6}{\eta^6}  
 \left( 
 F^{(1, 0)}( \tau, \tfrac{ \tau}{2} )  + (-)^m F^{(1, 1)} ( \tau, \tfrac{ \tau}{2} ) \right)
 \right\}  \right]. 
\end{align}
Though the above expression is lengthy, the structure of the index is quite easy to decipher. 
To see this, let us list the characters of the the $D_6$ lattice. Consider the lattice in the  fermionic 
representation.   Then we have the  following partition functions for the various sectors. 
\begin{eqnarray}
 {\cal Z} ( D6; NS^+; q ) = \frac{\theta_3^6}{\eta^6} , \quad
 {\cal Z} ( D6; NS^- , R; q) = \frac{ \theta_4^6}{\eta^6}, \quad
 {\cal Z} ( D6; R; q ) = \frac{ \theta_2^6}{\eta^6} . 
\end{eqnarray}
Here $NS^- $ refers to the Neveu-Schwarz sector with $(-1)^F$ inserted in the trace. $F$ is the 
worldsheet fermion number of these left moving fermions of the $D_6$ lattice. $R $ refers to the 
Ramond  sector. 
From (\ref{fullans1}) we note that the 
 coefficients  of these $D_6$ partitions functions are  the twisted elliptic genus of $\mathbb{Z}_2$
 CHL orbifold of $K3$. The contribution of ${\cal Z} ( D6; NS^- , R; q)$ is weighted
 with 
 $-1$.   It is important to note that  the new supersymmetric index given in (\ref{fullans})  
 was obtained by an explict calculation and it admitted a decomposition  in the form given 
 in (\ref{fullans1}). It is interesting that  the structure 
 seen for $K3\times T^2$ by \cite{Harvey-Moore,Cardoso-Curio-Lust}  in which the elliptic genus of the 
 internal CFT plays the role in determining 
 the new supersymmetric index is generalized
 to the twisted elliptic genus for the CHL compactification.

\subsubsection*{Decomposition in terms of Eisenstein series}

It is also useful to rewrite the new supersymmetric index in (\ref{fullans}) in another form 
to obtain the gauge threshold corrections. 
For this,  note that we have the following identities between modular forms. 
\begin{align}\label{mident}
-(\th_3^8 \th_4^4 + \th_4^8 \th_3^4 ) &= -\frac{2}{3} \left(E_6 + 2 \E_2(\tau) E_4 \right),  \\ \nonumber
\th_3^8 \th_2^4 + \th_2^8 \th_3^4  &= -\frac{2}{3} \left(E_6 - \E_2\left(\tfrac{\tau}{2}\right) E_4 \right),  \\ \nonumber
\th_2^8 \th_4^4 - \th_2^8 \th_4^4  &= -\frac{2}{3} \left(E_6 - \E_2\left(\tfrac{\tau+1}{2}\right) E_4 \right), 
\end{align}
where \footnote{The  modular function  $\E_N$  was introduced in \cite{justin1} where it was called $E_N$} 
\begin{eqnarray}
\E_N (\tau) = \frac{12 i}{\pi (N-1)} \pd_\tau \log \frac{\eta(\tau)}{\eta(N\tau)}.
\end{eqnarray}
The identities in   (\ref{mident}) have  been verified 
by performing a $q$-expansion which is detailed in the appendix \ref{app:a}. 
Substituting these identities in (\ref{fullans}) we obtain the
form 
\begin{eqnarray}\label{fullans2}
 {\cal Z}_{ {\rm new} }^{(2)}  ( q , \bar q ) = - \frac{2 E_4}{\eta^{12} }\times \left[   \sum_{ {m_1, m_2, n_1 n_2 \in \mathbb{Z }}}
 q^{\frac{p_L^2}{2} } \bar q ^{\frac{p_R^2}{2}}   \frac{1}{\eta^{12}} 
 \left\{   2  E_6  +  ( -1)^m\frac{2}{3}
 \left(  E_6 + 2 \E_2(\tau)   E_4 \right)  \right\}  \right.   \nonumber  \\ 
 \nonumber
   \left.  \qquad 
 +\sum_{ {\substack{ {m_1, m_2, n_2 \in  \, \mathbb{Z }}, \\ {n_1 \in \, \mathbb{Z} + \frac{1}{2}} } } } q^{\frac{p_L^2}{2} } \bar q ^{\frac{p_R^2}{2}}
 \frac{2 }{3\eta^{12}} \left\{   \left( E_6  - \E_2( \tfrac{\tau}{2} ) E_4 \right)     	+ (-1)^m   \left( E_ 6 - \E_2 ( \tfrac{\tau+1}{2} ) E_4 \right)  
  \right\} 
 \right]. \\
\end{eqnarray}
It is also instructive to derive the the  expression  in (\ref{fullans2}) for the new supersymmetric index directly from 
from  (\ref{fullans1}).  For this we use the more general form for the twisted elliptic genus of the $N=2$ CHL 
orbifold of $K3$ from \cite{justin1}.
\begin{eqnarray}\label{eillipchl2}
 F^{(0, 0)} ( \tau, z)  &=& 4A(\tau, z) ,  \qquad F^{(0, 1)} ( \tau, z)  = \frac{4}{3} A(\tau, z) - \frac{2}{3} B(\tau, z) \E(\tau) , 
 \\  \nonumber
 F^{(1, 0) } ( \tau, z) &=& \frac{4}{3} A(\tau, z) + \frac{1}{3} B(\tau, z) \E_2(\tfrac{\tau}{2} ), \qquad
 F^{(1, 1) } (\tau, z) =  \frac{4}{3} A(\tau, z) + \frac{1}{3} B(\tau, z) \E_2(\tfrac{\tau+ 1}{2} ), 
\end{eqnarray}
where 
\begin{eqnarray}\label{defab}
 A(\tau, z) = \frac{\theta_2 (\tau, z)^2 }{\theta_2 ( \tau, 0)^2} + 
 \frac{\theta_3 (\tau, z)^2 }{\theta_3 ( \tau, 0)^2}+
 \frac{\theta_4 (\tau, z)^2 }{\theta_4 ( \tau, 0)^2} , \qquad 
 B(\tau, z) = \frac{\theta_1(\tau, z)^2}{\eta^6}. 
\end{eqnarray}
Substituting these forms for the twisted Elliptic genus in (\ref{fullans1}) it is easy to see that it organizes into 
the form (\ref{fullans2}).   To show this it is convenient to use the identities
\begin{align} \label{abrel}
 A(\tau, \tfrac{1}{2}  ) &=  \frac{\left( \theta_4^4 \theta_2^2 +  \theta_3^4 \theta_2^2  \right)}{4 \eta^6},     
 \hspace*{2.5cm} B(\tau, \tfrac{1}{2} ) = \frac{\theta_2^2}{ \eta^6},  \\ \nonumber
 A(\tau, \tfrac{\tau}{2}) &=  \frac{q^{-1/4} \left( 
    \theta_3^4 \theta_4^2 +  \theta_2^4 \theta_4^2 \right) }{4 \eta^6} , \qquad 
\ \ \ \ \   B(\tau, \tfrac{\tau}{2} ) = -  \frac{q^{-1/4}  \theta_4^2}{ \eta^6}, \\ \nonumber
 A(\tau, \tfrac{\tau +1}{2} ) &= \frac{q^{-1/4} \left( -
    \theta_4^4 \theta_3^2 +  \theta_2^4 \theta_3^2 \right) }{4 \eta^6}, \qquad 
   B(\tau, \tfrac{\tau +1}{2} ) =   \frac{q^{-1/4} \theta_3^2}{ \eta^6}. 
\end{align}
Using these identities in (\ref{fullans1}) we obtain (\ref{fullans2}).

\subsubsection*{Modular invariance }

The new supersymmetric index has the property that 
$\tau_2 {\cal Z}_{ {\rm new}} (\tau, \bar \tau)$  has to be an  $SL(2, \mathbb{Z})  $ non-holomorphic  modular form 
of weight $-2$. This is essentially because it occurs in threshold integrals  along with 
modular forms of weight $2$ \footnote{This will be seen in section \ref{sec:5}.}
 and the integrand in any threshold integral  has to  be modular invariant. 
Let us now verify that  $\tau_2 {\cal Z}_{ {\rm new}} $ indeed transforms as a weight $-2$ 
modular form. For this, we need the following transformation property of 
$\E_N$ 
\begin{equation}\label{transen}
 \E_N(\tau + 1) = \E_N(\tau) , \qquad  \E_N( -1/ \tau ) =   - \tau^2 \frac{1}{N} \E_N ( \tau/N ) .
\end{equation}
Using this property, it is easy to see that for the special case of $N=2$ we have
\begin{equation}\label{transe2}
 \E_2(- \tfrac{1}{2\tau} ) = - 2 \tau^2 \E_2(  \tau ) , \qquad 
 \E_2( - \tfrac{1}{2\tau} + \tfrac{1}{2} ) = \tau^2 \E_2( \tfrac{\tau +1}{2} ) .
\end{equation}
Let us define the following lattice sums  over $T^2$ 
\begin{eqnarray} \label{lsums}
 \Gamma_{2, 2}^{(0, 0)}  ( \tau, \bar \tau) &=&  \sum_{m_1, m_2, n_1 , n_2 \in \mathbb{Z} }  q^{\frac{p_L^2}{2} } 
 \bar  q^{\frac{p_R^2}{2} }, \\ \nonumber
 \Gamma_{2,2}^{(0, 1) } ( \tau, \bar \tau) &=& \sum_{m_1, m_2, n_1 , n_2 \in \mathbb{Z} } q^{\frac{p_L^2}{2} } 
 \bar  q^{\frac{p_R^2}{2} } (-1)^{m_1}, \\ \nonumber
 \Gamma_{2,2}^{(1, 0) } ( \tau, \bar \tau) &=& 
   \sum_{ {\substack{ {m_1, m_2, n_2 \in  \, \mathbb{Z }}, \\ {n_1 \in \, \mathbb{Z} + \frac{1}{2}} } } }
 q^{\frac{p_L^2}{2} } 
 \bar  q^{\frac{p_R^2}{2} }, \\ \nonumber
 \Gamma_{2,2}^{(1, 1) } ( \tau, \bar \tau) &=& 
 \sum_{\substack{{m_1, m_2, n_2 \in \mathbb{Z},}\\ { n_1 \in \mathbb{Z} + \frac{1}{2} }}} \tilde q^{\frac{p_L^2}{2} } 
 \bar  q^{\frac{p_R^2}{2} }(-1)^{m_1} . 
\end{eqnarray}
From the expression for $p_L, P_R$ given in (\ref{plpr}) it is easy to see that under the shift $\tau \rightarrow \tau+1$, 
we obtain the following relations between the lattice sums
\begin{eqnarray} \label{ttrans}
 \tau_2\Gamma_{2, 2}^{(0,0)} ( \tau +1, \bar \tau + 1) &=&\tau_2 \Gamma_{2,2}^{(0,0)} ( \tau, \bar\tau), \\ \nonumber
 \tau_2  \Gamma_{2, 2}^{(0,1)} ( \tau +1, \bar \tau + 1) &=& \tau_2  \Gamma_{2,2}^{(0,1)} ( \tau, \bar\tau), 
  \\ \nonumber
  \tau_2 \Gamma_{2, 2}^{(1,0)} ( \tau +1, \bar \tau + 1) &=&  \tau_2 \Gamma_{2,2}^{(1,1)} ( \tau, \bar\tau), \\ \nonumber
  \tau_2 \Gamma_{2, 2}^{(1,1)} ( \tau +1, \bar \tau + 1) &=&  \tau_2 \Gamma_{2,2}^{(1,0)} ( \tau, \bar\tau).
\end{eqnarray}
Using Poisson resummation one can show that under the transformation $\tau \rightarrow - 1/\tau$ 
the following relations hold
\begin{eqnarray}\label{strans}
 \left( -1/\tau\right)_2  \Gamma_{2, 2}^{(0,0)}( -{1}/{\tau} , -{1}/{\bar\tau} ) 
  &=& \tau_2 \Gamma_{2,2}^{(0,0)} ( \tau, \bar\tau), \\ \nonumber
 \left( -1/\tau\right)_2  \Gamma_{2, 2}^{(0,1)}( -{1}/{\tau} , -{1}/{\bar\tau} ) 
  &=& \tau_2 \Gamma_{2,2}^{(1,0)} ( \tau, \bar\tau), \\ \nonumber
 \left( -1/\tau\right)_2   \Gamma_{2, 2}^{(1,0)}( -{1}/{\tau} , -{1}/{\bar\tau} ) 
  &=& \tau_2 \Gamma_{2,2}^{(0,1)} ( \tau, \bar\tau), \\ \nonumber
 \left( -1/\tau\right)_2   \Gamma_{2, 2}^{(1,1)}( -{1}/{\tau} , -{1}/{\bar\tau} ) 
  &=& \tau_2 \Gamma_{2,2}^{(1,1)} ( \tau, \bar\tau).
\end{eqnarray}
Using the equations (\ref{transen}), (\ref{transe2}), (\ref{ttrans}) and (\ref{strans}) it is 
easy to see that  $\tau_2 {\cal Z}_{ {\rm new}}^{(2)}$ where the 
new supersymmetric index given in the form (\ref{fullans2}) is a modular form of 
weight $-2$. To demonstrate this we have to also  use the fact that 
$\eta,  E_4, E_6$ are modular forms of weight $1/2, 4, 6$ respectively. 
This result ensures that the result for the integrand in the threshold corrections is modular invariant.

\subsection{The $\mathbb{Z}_N$ orbifold}

From the explicit calculation and the discussions in the earlier section for the $N=2$ CHL orbifold of $K3$ it is 
easy to arrive at the expression for the new supersymmetric index for the other values of $N$. 
To write down the  expression for the index it is useful to define the following 
\begin{eqnarray}
 {\cal I}^{(r , s) }_{RR}(q, \bar q) &=&  \sum_{\substack{m_1, m_2, n_2 \in \mathbb{Z}, \\ n_1 = \mathbb{Z} + \frac{r}{N} }}
 q^{ \frac{p_L^2}{2} } \bar q^{\frac{p_R^2}{2}}  e^{2\pi i m_1 s/N} 
 F^{(r, s)} ( \tau, \tfrac{1}{2} ) , \\ \nonumber
 {\cal I}^{(r , s) }_{(NS^+)}(q, \bar q) &=&  q^{1/4} \sum_{\substack{m_1, m_2, n_2 \in \mathbb{Z}, \\ n_1 = \mathbb{Z} + \tfrac{r}{N} }}
 q^{ \frac{p_L^2}{2} } \bar q^{\frac{p_R^2}{2}}  e^{2\pi i m_1 s/N} 
 F^{(r, s)} ( \tau, \tfrac{\tau +1}{2} ), \\ \nonumber
 {\cal I}^{(r , s) }_{(NS^-)}(q, \bar q) &=&
 - q^{1/4}  \sum_{\substack{m_1, m_2, n_2 \in \mathbb{Z}, \\  n_1 = \mathbb{Z} + \frac{r}{N} }}
 q^{ \frac{p_L^2}{2} } \bar q^{\frac{p_R^2}{2}}  e^{2\pi i m_1 s/N} 
 F^{(r, s)} ( \tau, \tfrac{\tau }{2} ), \\ \nonumber
 & & \qquad\qquad\qquad \qquad \qquad\qquad {\hbox{for}} \; 0\leq r, s, \leq N-1. 
\end{eqnarray}
Here $F^{(r, s)} (\tau, z )$ is the twisted elliptic genus of the  
CHL orbifold of $K3$ which is given by  \cite{justin1}
\begin{eqnarray}\label{ntwist}
 F^{(0, 0)} ( \tau, z) &=& \frac{8}{N} A(\tau, z) , \\ \nonumber
 F^{(0, s)} ( \tau, z) &=& \frac{8}{N( N+1) } A(\tau, z) 
 - \frac{2}{N+1}  \E_N(\tau)B(\tau, z)   , \quad {\hbox{for }}  1\leq s \leq N-1, 
 \\  \nonumber
 F^{(r, rk)} (\tau, z) &=&  \frac{8}{N(N+1) } A(\tau, z) + \frac{2}{N( N+1)} \E_N \left( \tfrac{\tau +k}{N} \right) B(\tau, z) , \\
 \nonumber
 & &  \qquad\qquad\qquad\qquad {\hbox{for }} \; 1\leq r \leq N-1, \;  0\leq k \leq N-1, 
\end{eqnarray}
where $A(\tau, z), B(\tau, z) $ are defined in (\ref{defab}).  Using these definitions, the new supersymmetric 
index for the $\mathbb{Z}_N$ CHL orbifold of $K3$ is given by 
\begin{eqnarray}
 {\cal Z}^{(N)}_{{\rm new}} (q, \bar q) = \frac{2E_4}{\eta^{12}} \sum_{r, s =0}^{N-1}  \left[
 \frac{\theta_2^6}{\eta^6} {\cal I }^{(r,s)}_{R} 
 + \frac{\theta_3^6}{\eta^6} {\cal I }^{(r,s)}_{NS^+} 
 + \frac{\theta_4^6}{\eta^6} {\cal I }^{(r,s)}_{NS^-}  \right] . 
\end{eqnarray}
Substituting  the expressions for the twisted elliptic genus from (\ref{ntwist})  and using the 
relations in (\ref{abrel}) we obtain the following expression for the new supersymmetric index in terms
of Eisenstein functions
\begin{eqnarray}\label{fullans3}
  & &{\cal Z}^{(N)}_{{\rm new}} (q, \bar q) =  -  \frac{2E_4}{\eta^{12}} \times \\ \nonumber
 & &  \left[ \sum_{m_1, m_2, n_2 n_2 \in \mathbb{Z} } q^{\frac{p_L^2}{2} } \bar q^{ \frac{p_R^2}{2}} 
  \frac{1}{\eta^{12}} \left \{  \frac{4}{N}  E_6  + \left( \sum_{s =1}^{N-1} e^{\frac{2\pi i s m_1}{N} } \right) 
  \left( \frac{4}{N(N+1) } E_6  + \frac{4}{N+1} \E_N(\tau)  E_4  \right) \right\} \right. 
  \\ \nonumber
  & & \left. + \sum_{r= 1}^{N-1}  \sum_{\substack{{m_1, m_2, n_2 \in \mathbb{Z},}\\{ n_1 = \mathbb{Z} + \frac{r}{N} }}}
   q^{\frac{p_L^2}{2} } \bar q^{ \frac{p_R^2}{2}} 
   \sum_{ k = 0 }^{N-1} \frac{ e^{ \frac{2\pi i rk  m_1}{N} } }{\eta^{12}}
  \left\{  
  \frac{4}{N( N+1) } E_6  - \frac{2}{N( N+1) } \E_N( \frac{\tau + k}{N} ) E_4  \right\}  \right]. 
\end{eqnarray}
A simple check of the above formula is that it reduces to (\ref{fullans2}) for the $N=2$ case. 
One can re-write this expression by performing the sum over the phases wherever possible, but 
it is convenient to keep the expression as it is. 
It can be shown  that $\tau_2 {\cal Z}^{(N)}_{{\rm new}} (q, \bar q)$ 
is a modular form of weight $-2$  by generalizing the  method discussed for the $N=2$ case in 
detail.  Therefore the structure of the new elliptic index for CHL orbifolds  of $K3$ is such that 
the Eisenstein function $E_6$ which occurs for the $K3$ is modified to the 
form given in the curly brackets of the expression in (\ref{fullans3}).

\section{Mathieu moonshine}\label{sec:4}

From the analysis of the new supersymmetric index for CHL orbifolds of $K3$ we have seen that
it is essentially determined by the twisted elliptic index of  $K3$. 
This  property is essentially seen in the expressions 
(\ref{fullans2}) for the $N=2$  orbifold and (\ref{fullans3}) for  other values of $N$. 
It is known \cite{Cheng:2010pq,Gaberdiel:2010ch,Gaberdiel:2010ca,Eguchi:2010fg} that the twisted elliptic genus of $K3$ admits $M_{24}$ symmetry. 
Therefore, it must be possible to discover the $M_{24}$ representations 
in the new supersymmetric index for the CHL orbifolds of $K3$, just as it
was done for the new supersymmetric index for $K3$ compactifications in \cite{Kachru}. 

Let us first recall how Mathieu moonshine  -- i.e.~$M_{24}$ representations --  is seen in the elliptic genus of $K3$. It is given by 
\begin{equation}
Z_{K3} (\tau, z) = 8 A(\tau, z).
\end{equation} 
 Let us decompose the elliptic genus into the elliptic genera of the 
short and the long representations of the 
${\cal N}=4$  super conformal algebra.  These are given by \cite{Eguchi:1988vra}
\begin{eqnarray}
{\rm{ch}}_{h =  \frac{1}{4},  l = 0 } (\tau, z )   &=&  - i e^{\pi i z} \frac{\theta_1( \tau, z ) }{\eta (\tau)^3} 
\sum_{n = - \infty}^{\infty} \frac{ 1}{1 - e^{2\pi i ( n\tau  + z) } }
 e^{\pi  i\tau n ( n + 1) } e^{ 2\pi i ( n  + \frac{1}{2} ) }  , \\ \nonumber
 {\rm{ch}}_{h =  n+ \frac{1}{4},  l =  \frac{1}{2}  } (\tau, z )  & = & e^{2\pi i \tau ( n  - \frac{1}{8} )  } 
 \frac{\theta_1( \tau , z) ^2}{\eta(\tau) ^2}. 
\end{eqnarray}
Then we have 
\begin{equation}
Z_{K3}(\tau, z ) = 24 {\rm{ch}}_{h =  \frac{1}{4},  l = 0 } (\tau, z )  + 
\sum_{n=0}^\infty  A_n^{(1)} {\rm{ch}}_{h =  n+ \frac{1}{4},  l =  \frac{1}{2}  } (\tau, z ).
\end{equation}
where the first few values of $A_n^{(1)}$ are given by 
\begin{equation}
A_n^{(1)} = -2,\,  90,\,  462,\,  1540,\,  4554,\,  11592, \ldots
\end{equation}
These coefficients are  either the  dimensions  or the sums of dimensions 
of the irreducible representations of the group $M_{24}$ \cite{Eguchi:2010ej}. 
The generalization of this observation  to the twisted elliptic genus 
of $K3$ was done in \cite{Cheng:2010pq,Gaberdiel:2010ch,Eguchi:2010fg}. 
Let us first discuss the $N=2$ CHL orbifold of $K3$. 
Consider the twisted elliptic index 
\begin{equation}
 2 F^{(0,1)}( \tau, z) = \frac{8}{3} A(\tau, z) -\frac{4}{3} B(\tau, z) \E_2 (\tau). 
\end{equation}
This admits the following decomposition in terms of ${\cal N}=4$ Virasoro characters
\begin{eqnarray} \label{dellip}
  2 F^{(0,1)}( \tau, z)  = 8 {\rm{ch}}_{h =  \frac{1}{4},  l = 0 } (\tau, z )  +
  \sum_{n=0}^\infty  A_n^{(2)} {\rm{ch}}_{h =  n+ \frac{1}{4},  l =  \frac{1}{2}  } (\tau, z ).
\end{eqnarray}
Where the coefficient $8$ is the twisted Euler number  
of $K$ which is given by 
\begin{equation}
 \chi_N = \frac{24}{N+1}, \qquad N = 2, 3, 5, 7 .
\end{equation}
In (\ref{dellip})   the first few values of $A_n^{(2)}$ are given by 
\begin{equation} \label{mackay2}
 A_n^{(2)} = -2,  \, -6 , \, 14 , \, -28 , \, 42 , \, -56 , \, 86 , \, -138 , \,  \ldots
\end{equation}
These coefficients can be identified with McKay-Thompson series constructed out of trace of 
the element $g$ corresponding to the $\mathbb{Z}_2$ involution of $K3$  embedded in $M_{24}$. 
From the structure of the  new supersymmetric index in (\ref{fullans1}) and (\ref{fullans2}) 
 the  new supersymmetric index in the $(0,1)$ sector given by 
\begin{eqnarray} \label{defg2}
G^{(2)} (q)  = - \frac{4}{3} \left[   \frac{E_6 + 2 \E_2(\tau)E_4}{\eta^{12}}   \right] .
\end{eqnarray}
We have multiplied by a factor of $2$ to agree with the normalizations of the twisted 
elliptic genus of $K3$ used in \cite{Cheng:2010pq}. 
Then the new supersymmetric index in the $(0,1)$ sector 
admits the following decomposition
\begin{equation}\label{defg22}
 G^{(2)}(q) =  8 {g}_{h =  \frac{1}{4},  l = 0 } (\tau )  +
  \sum_{n=0}^\infty  A_n^{(2)} {g}_{h =  n+ \frac{1}{4},  l =  \frac{1}{2}  } (\tau),
\end{equation}
where 
\begin{eqnarray} \nonumber
 g_{h =  \frac{1}{4},  l = 0 } (\tau)   =  \frac{\theta_2^6}{\eta^6} {\rm{ch}}_{h =  \frac{1}{4},  l = 0 } (\tau, \frac{1}{2} )
 + q^{1/4}\frac{\theta_3^6}{\eta^6} {\rm{ch}}_{h =  \frac{1}{4},  l = 0 } (\tau, \frac{ 1+ \tau }{2} )
 - q^{1/4} \frac{\theta_4^6}{\eta^6} {\rm{ch}}_{h =  \frac{1}{4},  l = 0 } (\tau, \frac{\tau}{2} ), 
 \\ \nonumber
  g_{h =  \frac{1}{4},  l = 0 } (\tau)   =  \frac{\theta_2^6}{\eta^6} {\rm{ch}}_{h =  \frac{1}{4},  l = \frac{l}{2}} (\tau, \frac{1}{2} )
 + q^{1/4}\frac{\theta_3^6}{\eta^6} {\rm{ch}}_{h =  \frac{1}{4},  l = 0 } (\tau, \frac{ 1+ \tau }{2} )
 - q^{1/4}\frac{\theta_4^6}{\eta^6} {\rm{ch}}_{h =  \frac{1}{4},  l = 0 } (\tau, \frac{\tau}{2} ).  \\ \label{defgs}
\end{eqnarray}
The $g$'s are products of characters of $D_6$ and ${\cal N}=4$ Virasoro characters. 
$G^{(2)}$ given in (\ref{defg2}) is the generalization of 
\begin{equation}
 G^{(1)}(q) = - 2 \frac{E_6}{\eta^{12}}
\end{equation}
which is the new supersymmetric index for $K3$ compactifications. 
Substituting the expressions for $g$'s from  (\ref{defgs}) into  (\ref{defg22})  and using (\ref{defg2}) we can 
solve for the coefficients $A_n^{(2)}$. 
We have checked using Mathematica 
that the first 8  coefficients fall into the McKay-Thompson series for the 
$\mathbb{Z}_2$ involution embedded in $M_{24}$ given in (\ref{mackay2}).

Let us now proceed with the analysis for other values of $N$.  From (\ref{fullans3}) we see that 
 the  new supersymmetric index in the $(0,1)$ sector is given by 
\begin{equation}\label{defgnn}
 G^{(N)} (q) =\frac{ -N }{\eta^{12}}{ \left[  \frac{4}{N(N+1)} E_6 +\frac{4}{N+1} \E_N(\tau) E_4  \right] } .
\end{equation}
Here we have multiplied a factor of $N$ to agree with the normalizations of
the twisted elliptic genus of $K3$ in \cite{Cheng:2010pq}. 
Let us write $G^{(N)}$ as 
\begin{equation} \label{defgn}
 G^{(N)}(q) =  \chi_N {g}_{h =  \frac{1}{4},  l = 0 } (\tau )  +
  \sum_{n=0}^\infty  A_n^{(N)} {g}_{h =  n+ \frac{1}{4},  l =  \frac{1}{2}  } (\tau).
\end{equation}
By equating  (\ref{defgn}) and (\ref{defgnn}) we can solve for  the coefficients $A_n^{(N)}$
\footnote{A Mathematica routine was used for this.}. 
The first few coefficients are given by 
\begin{align}
A^{(3)}_n &= -2,  \, 0 , \, -6 , \, 10 , \, 0 , \, -18 , \, 20 , \, 0 , \,  \dots,  \nn \\
A^{(5)}_n &= -2, \, 0,\,  2,\,   0, \,  -6,  \, 2, \,   0, \, 6\, , \dots,  \nn \\
A^{(7)}_n &=  -2, \,-1, \,0, \,
0, \,4, \, 0, \, -2, \, 2 , \dots. 
\end{align}
As expected these are the coefficients of the McKay-Thompson series for the $\mathbb{Z}_N$ 
involution of $K3$ embedded in $M_{24}$ \footnote{Compare table 1 of \cite{Cheng:2010pq}.} . 
This analysis serves as a consistency check for the 
the new supersymmetric index for the $\mathbb{Z}_N$ orbifolds of $K3$. 
The analysis also indicates that the BPS states in these compactifications 
have a decomposition in terms of the coefficients of the McKay-Thompson series. 

As we have seen explicitly, for the $N=2$ case, the new supersymmetric index in the  $(1,0)$ twisted 
sector is related to that of the $(0,1)$ sector by the modular transformation $\tau \rightarrow - 1/\tau$. 
This is also true for other values of $N$. 
This implies that the new supersymmetric index in these sectors must also contain the 
a modular transformed version of the McKay-Thompson series. 
It will be interesting to show this explicitly. 
There are 26 McKay-Thompson series corresponding to the $26$ conjugacy classes of 
$M_{24}$. It will be interesting to 
 to construct  and study the properties of the  
the new supersymmetric index corresponding to remaining classes. 
The twisted elliptic genera of $K3$  for each of these classes have been constructed in 
\cite{Cheng:2010pq,Gaberdiel:2010ch,Gaberdiel:2010ca,Eguchi:2010fg} \footnote{See \cite{Govindarajan:2009qt} for an 
earlier  explicit construction of the twisted elliptic genus for  the $N=4$ orbifold. } which will be a good starting point for this study.

 \def\no{n^{(1)}}
 \def\nt{n^{(2)}}
\section{Gauge threshold corrections}\label{sec:5}

In this section, we will evaluate the one-loop threshold corrections for each of the two unbroken gauge groups
 $E_7$ and $E_8$ as a function of the K\"{a}hler and complex structure moduli  and  the
 Wilson line modulus  on $T^2$ for the  heterotic compactifications on CHL orbifolds of $K3$. 
 To begin we will recall the evaluation of the threshold integrals for the gauge couplings 
 of heterotic on $K3 \times T^2$. We then proceed to generalize to the case 
 of the $\mathbb{Z}_2$ CHL orbifold and then present the results for the $\mathbb{Z}_N$
 orbifold with $N=3, 5, 7$. 
 We will show that the difference in the threshold integrals of the two unbroken gauge groups
 reduces to Siegel modular forms associated with dyon partition functions in ${\cal N}=4$ string 
 compactifications studied in \cite{justin1}. 
 
 \subsection{Thresholds in $K3\times T^2$}
 
 Let us first discuss the situation without the Wilson line turned on. 
 The  moduli dependence of the one-loop running of the gauge group is given by 
 \begin{equation}\label{thresh}
  \Delta_{G} (T, U) = \int_{\cal F}  \frac{d^2 \tau}{\tau_2} \left( {\cal B}_G - b (G)  \right) , 
 \end{equation}
 where ${\cal B}$ is a trace over the internal Hilbert space which is defined as
 \begin{equation}\label{defbg}
 {\cal B}_G(\tau, \bar \tau)  = \frac{1}{\eta^2} {\rm Tr}_R\left\{ Fe^{i\pi F } q^{L_0  - \frac{c}{24} } \tilde 
 \bar q^{\tilde L_0  - \frac{\tilde c}{24}}  \left(  Q^2 ( G) - \frac{1}{8\pi \tau_2} \right)  \right\}, 
 \end{equation}
where $Q$ is the charge of the lattice vectors. 
The coefficient $b(G)$ is the one-loop beta function which is present to ensure that the integral is 
well-defined  in the limit $\tau_2\rightarrow \infty$. 
Since we will be interested only in the moduli dependence, this coefficient  will not 
play a crucial role in our analysis. 
Note that ${\cal B}$ is closely related to the 
new supersymmetric index. In fact the term proportional to $1/8\pi \tau_2$ is the  new supersymmetric
index.  The easiest way to determine the  term with the charge insertion $Q^2(G) $ is to consider 
the action of $q\partial_q$ on the  partition function of the 
appropriate lattice sum so that  $\tau_2 {\cal B}$ is 
modular invariant. 
The integral in (\ref{thresh}) is carried out over the fundamental domain. 

Let us recall how to evaluate  the one-loop threshold integrands for the groups $E_7$ and $E_8$ 
for the $K3\times T^2$ compactifications. 
For group $E_8$, the integrand is given by 
\begin{equation}\label{bexp}
{\cal B}_{E_8}^{(1)} (\tau \bar \tau) = -2 \Gamma_{2, 2} (q, \bar q )\frac{1}{\eta^{24}} 
\left( \alpha_G q\partial_q E_4  - \frac{1}{8\pi \tau_2}   \right)  4  E_6 .
\end{equation}
Here we have supressed the moduli dependence of ${\cal B}$ which arises due to the 
lattice sum on $T^2$ given by $\Gamma_{2, 2}$.
Note that, this is essentially an   operation on the new supersymmetric index  for these 
compactifications which is given in (\ref{fact1}). 
The charge insertion of the $E_8$ lattice is obtained by 
the action of $q\partial_q$ on  the lattice sum $E_4(q) $. 
The coefficient $\alpha_G$ is determined by demanding 
$\tau_2 {\cal B}$ is modular invariant. 
To determine this coefficient consider the following identity due to 
Ramanujan
\begin{equation}\label{ram1}
 q\partial_q E_4 = \frac{1}{3}  ( E_2 E_4 - E_6). 
\end{equation}
Substituting this identity in (\ref{bexp}) we obtain 
\begin{equation}
 {\cal B}_{E_8}^{(1)} (\tau \bar \tau) = - 8 \Gamma_{2, 2} (q, \bar q )\frac{1}{\eta^{24} }  \left\{ 
 \left( \frac{\alpha_{G}} {3}  E_2 -  \frac{1}{8\pi \tau_2} \right) E_4  E_6 -  \frac{\alpha_{G }}{3}   E_6^2
 \right\}.
\end{equation}
It is now clear that choosing $\alpha_G= \frac{1}{8}$ ensures the the quasi-modular form $E_2$ 
occurs in the combination 
\begin{equation}\label{te2}
 \tilde E_2 = E_2 - \frac{3}{\pi \tau_2}.
\end{equation}
which transforms as a  good modular form of weight $2$. 
Therefore the threshold integrand for the gauge group $E_8$ is given by 
\begin{equation} \label{be8}
  {\cal B}_{E_8}^{(1)} (\tau , \bar \tau) = - \frac{1}{3}  \Gamma_{2, 2} (q, \bar q )\frac{1}{\eta^{24} }\left\{   \left( 
  E_2 - \frac{3}{\pi \tau_2} \right) E_4 E_6  - E_6^2   \right\}.
\end{equation}
Similarly the threshold integrand for the group $E_7$ is obtained  by evaluating  
\begin{equation}\label{bexp2}
{\cal B}_{E_7}^{(1)} (\tau,  \bar \tau) = -8 \Gamma_{2, 2} (q, \bar q )\frac{1}{\eta^{24}} 
\left( \alpha_{G'} q\partial_q E_6  - \frac{1}{8\pi \tau_2}   \right)   E_4 . 
\end{equation}
Now we have the Ramanujan identity 
\begin{equation}
  q\partial_q E_6 = \frac{1}{2}  ( E_2 E_6 - E_4^2) . 
\end{equation}
This identity together with modular invariance determines $\alpha_{G'} = 1/{12}$ . 
Thus the threshold integrand for the gauge group $E_7$ is given by 
\begin{equation}\label{be7} 
 {\cal B}_{E_7}^{(1)} ( \tau, \bar \tau) = - \frac{1}{3} \Gamma_{2, 2} (q, \bar q )\frac{1}{\eta^{24}}\left\{   \left(
 E_2 - \frac{3}{\pi \tau_2} \right) E_ 4 E_6 - E_4^3 \right\}. 
\end{equation}
Finally consider the difference in the threshold integrands for  the gauge groups in (\ref{be8}) and (\ref{be7}). 
We obtain 
\begin{eqnarray}
 {\cal B}_{E_7}^{(1)} - {\cal B} _{E_8}^{(1)} &=& \frac{1}{ 3 \eta^{24}}  \Gamma_{2, 2} \left( 
 E_4^3 - E_6^2 \right) , \\ \nonumber
 &=& 576  \Gamma_{2, 2} . 
\end{eqnarray}
To obtain the second line we have used the identity 
\begin{equation}
 E_4^3 - E_6^2 = 1728 \eta^{24} . 
\end{equation}
Therefore the threshold integral reduces to the trivial 
integral over the fundamental domain of just  the lattice sum  which is given by 
\begin{equation}
 \Delta_{E_7}^{(1)} ( T, U) - \Delta_{E_8}^{(1)} ( T, U) =  576 \int_{\cal F }  \frac{d^2 \tau}{\tau_2} (  \Gamma_{2,2}  -1)  .
\end{equation}
The constant $(-1)$  can be obtained by carefully keeping track of the constants $b_(G)$ in the 
threshold integrand (\ref{thresh}).  Essentially the $(-1)$ serves to regulate the integral as $\tau_2 \rightarrow \infty$. 
This integral was done by \cite{Dixon:1990pc} and the result reduces to the product of the 
Dedekind $\eta$ functions. 
\begin{equation}
\Delta_{E_7}^{(1)} ( T, U) - \Delta_{E_8}^{(1)} ( T, U) =- 48 \log ( T_2^{12} U_2^{12} |\eta(T) \eta( U) |^{48} ) .
\end{equation}
Here we are ignoring moduli independent constants. 
$T_2, U_2$ are the imaginary parts of the the $T, U$ moduli of the torus $T^2$.  
Note that the normalization of the thresholds used in this paper 
involves a division by the beta function compared to standard normalizations
in the literature. 
This is keep uniformity in the discussion when we evaluate the difference
in thresholds as well as when we turn to the CHL orbifolds.

\subsubsection*{Wilson line $V\neq 0$}

Let us now repeat this exercise with the Wilson line $V$  on the torus $T^2$ turned on. 
The Wilson line can be embedded either in the gauge group $E_8$ or $E_7$. 
We will take the Wilson line to be embedded in $E_8$ \footnote{ The discussion can be 
generalized when the Wilson line is embedded in $E_7$, with the same results. }. 
The procedure to evaluate gauge thresholds with the Wilson line was 
given in \cite{Cardoso-Curio-Lust}. 
Here we out line the steps. 
Due to the presence of the Wilson line, the lattice sum over $T^2$ is enhanced to 
$\Gamma_{3, 2}$ which is given by 
\begin{equation}
 \Gamma_{3, 2}=  \sum_{m_1, m_2, n_1, n_2, b} q^{\frac{p_L^2}{2}} \bar q^{\frac{p_R^2}{2}} . 
\end{equation}
where
\begin{eqnarray} \label{defprplv}
 \frac{p_R^2}{2} &=&  \frac{1}{4\, {\rm{det}}{\rm{Im}} \Omega }
 \left| -m_1 U + m_2  + n_1 T + n_2 ( TU - V^2) + b V \right|^2,  \\ \nonumber
 \frac{p_L^2}{2} &=&  \frac{p_R^2}{2} + m_1 n_1  + m_2 n_2  + \frac{1}{4} b^2 . 
\end{eqnarray}
and 
\begin{equation}
 \Omega = \left(  \begin{array}{cc}
                  U & V\\
                  V & T
                 \end{array}
\right) . 
\end{equation}
Thus the lattice sum over $T^2$ is characterized by the five charges
$(m_1, m_2, n_1, n_2, b)$. 
The new supersymmetric index with the Wilson line  is  then  determined by first re-writing 
 the lattice sum over $E_8$ in terms of a Jacobi form of index 1
 given by 
 \begin{equation}\label{defe41}
  E_{4, 1} ( \tau , z) = \frac{1}{2}\left[ 
  \theta_2( \tau, z ) ^2 \theta_2^6 + \theta_3(\tau, z )^2 \theta_3^6 + \theta_4( \tau, z)^2 \theta_4^6 \right] .
 \end{equation}
 Note that $E_{4, 1}(\tau , 0)  = E_4(q) $, essentially we have decomposed the $E_8$ lattice into  $D_6$ 
and $D_4$ and  introduced a chemical potential for the charges in the $D_4$ sub-lattice. 
This breaks the gauge group $E_8$ down to $SO(12) \times U(1)$ we will refer to this
group as $G$. 
We then decompose this Jacobi form  of index one into $SU(2)$ characters 
as follows
\begin{equation}
 E_{4, 1} (\tau, z) = E_{4, 1}^{\rm even}(q)  \theta_{\rm even} (\tau, z) + 
 E_{4, 1}^{\rm odd} ( q) \theta_{\rm odd} ( \tau, z)  .
\end{equation}
where 
\begin{eqnarray}
 \theta_{\rm even} (\tau, z) = \theta_3 ( 2\tau, 2z), \qquad\qquad \theta_{\rm odd} (\tau, z) = \theta_2 ( 2\tau, 2 z) .
\end{eqnarray}
This decomposition can be performed using the relations 
\begin{eqnarray} \label{thetasq}
 \theta_1^2( \tau, z) &=& \theta_2 ( 2\tau, 0) \theta_3 ( 2\tau, 2 z) - \theta_3 ( 2\tau, 0 ) \theta_2 ( 2\tau, 2 z) , \\ \nonumber
 \theta_2^2(\tau, z) &=& \theta_2 ( 2\tau, 0) \theta_3 ( 2\tau, 2 z) + \theta_3 ( 2\tau, 0 ) \theta_2 ( 2\tau, 2 z) , \\ \nonumber
  \theta_3^2(\tau, z) &=& \theta_3 ( 2\tau, 0) \theta_3 ( 2\tau, 2 z) + \theta_2 ( 2\tau, 0 ) \theta_2 ( 2\tau, 2 z), 
  \\ \nonumber
   \theta_4^2(\tau, z) &=& \theta_3 ( 2\tau, 0) \theta_3 ( 2\tau, 2 z) - \theta_2 ( 2\tau, 0 ) \theta_2 ( 2\tau, 2 z). 
\end{eqnarray}
Using these relations we get 
\begin{eqnarray}\label{e4evenodd}
 E_{4, 1}^{\rm even}(q) &=& \frac{1}{2} \left( 
 \theta_2 ( 2\tau, 0 ) \theta_2^6   + \theta_3 ( 2\tau, 0)  \theta_3^6 + \theta_3 ( 2\tau, 0 ) \theta_4^6 \right) , \\ \nonumber
 E_{4, 1}^{\rm odd} ( q) & =&  \frac{1}{2} \left( 
\theta_3 ( 2\tau, 0 ) \theta_2^6   + \theta_2 ( 2\tau, 0)  \theta_3^6 -  \theta_2 ( 2\tau, 0 ) \theta_4^6 \right) .
\end{eqnarray}
Note that the even and odd parts depend only on the modular parameter $\tau$. 
Finally the modified new supersymmetric index in the presence of the Wilson line is 
written as 
\begin{equation} \label{indexwil}
 {\cal Z}_{\rm new}^{(1)} ( q, \bar q) = - 8 \frac{E_6}{\eta^{24}} \left( 
 \sum_{ {\substack{{m_1, m_2, n_1, n_2 \in \mathbb{Z}, }\\ {b \in 2\mathbb{Z} } } }} q^{\frac{p_L^2}{2}} \bar q^{\frac{p_R^2}{2}} 
 E_{4, 1}^{\rm even}(q) + 
  \sum_{ {\substack{{m_1, m_2, n_1, n_2 \in \mathbb{Z}, }\\ {b \in 2\mathbb{Z}+1} } }} 
  q^{\frac{p_L^2}{2}} \bar q^{\frac{p_R^2}{2}}
  E_{4, 1}^ {\rm odd} ( q)  \right). 
\end{equation}
Here $p_L, p_R$ contain the K\"{a}hler,  complex structure  and the Wilson  line moduli dependence of the $T^2$.  
A similar procedure can be carried out when the Wilson line is embedded in the unbroken group $E_7$. 
In this situation the Jacobi form  $E_{6, 1}$ given by 
\begin{equation}\label{defe61}
 E_{6, 1}( \tau, z) = \frac{1}{2}\left(  - \theta_2^6 ( \theta_3^4 + \theta_4^4) \theta_2^2 ( \tau, z) 
 + \theta_3^6( \theta_4^4 - \theta_2^4) \theta_3^2 ( \tau, z)
 + \theta_4^6( \theta_2^4 + \theta_3^4) \theta_4^2 ( \tau, z) 
 \right)  .
\end{equation}
must be decomposed into its even and odd parts. 
The coupling of the lattice sum $\Gamma_{3, 2}$ to the even and odd parts of $E_{4, 1}$  in (\ref{indexwil}) 
is compactly denoted  as 
\begin{equation}
 {\cal Z}_{\rm new}^{(1)}  ( q, \bar q) = - 8\frac{E_6}{\eta^{24}} E_{4, 1} \otimes \Gamma_{3, 2} ( q, \bar q) . 
\end{equation}

Now we move to evaluating  the  integrand ${\cal B}_G$ in the gauge thresholds with the Wilson line. 
Let us evaluate the threshold integrand for the group $E_8$ first. 
To determine the coefficient of the $\alpha_{G}$ in the action  $q \partial_q$  we need the 
following identity analogous to (\ref{ram1})  which is given in \cite{Kawai:1996te,LopesCardoso:1996zj}
\begin{equation}\label{ram3}
 q\partial_q E_{4, 1}^{\rm{even, odd} } = \frac{7}{24} 
 \left( E_2 E_{4, 1}^{{\rm even, odd} } - E_{6, 1}^{\rm {even, odd}}  \right) . 
\end{equation}
For completeness we also provide the identity which is required if the Wilson line is embedded 
in $E_7$
\begin{equation}
 q\partial_q E_{6, 1}^{\rm{even, odd} } = \frac{11}{24} 
 \left( E_2 E_{6, 1}^{{\rm even, odd} } - E_{4, 1}^{\rm {even, odd}} E_4   \right) . 
 \end{equation}
From  (\ref{ram3}) 
it is easy to see that to preserve modular invariance we need $\alpha_{G} = 1/7$. 
Therefore we obtain 
\begin{equation}
 {\cal B}_{G}^{(1)}  ( \tau , \bar \tau ) =   
   - \frac{1}{3} \frac{1}{\eta^{24} }\left\{   \left( 
  E_2 - \frac{3}{\pi \tau_2} \right) E_{4, 1}  E_6  - E_{6, 1}E_6   \right\} \otimes \Gamma_{3, 2} ( q, \bar q ) .  
\end{equation}
The  threshold integrand  ${\cal B}_{G'}$ for group $E_7$  is given by 
\begin{equation}
 {\cal B}_{G'}^{(1)}  ( \tau , \bar \tau ) =   
   - \frac{1}{3} \frac{1}{\eta^{24} }\left\{   \left(
 E_2 - \frac{3}{\pi \tau_2} \right) E_ {4, 1} E_6 - E_4^2 E_{4, 1}  \right\} \otimes \Gamma_{3, 2} ( q, \bar q ). 
\end{equation}

Let us now take the difference between threshold corrections corresponding to the two gauge groups. 
We obtain 
\begin{equation}\label{integb}
 \Delta_{G' }^{(1)}( T, U, V) - \Delta_{G}^{(1)} ( T, U, V) = 
 \int_{\cal F}  \frac{d^2 \tau}{\tau_2 }  \frac{1}{3  \eta^{24}} 
 \left( E_4^2 E_{4, 1} - E_6 E_{6, 1}  \right) \otimes \Gamma_{3, 2} ( q, \bar q) .  
\end{equation}
Here we have ignored the constant term  in the integrand which can be determined 
by examining the behaviour of the integrand as $\tau_2 \rightarrow \infty$. 
The  combination  of the Eisenstein series which occurs in the (\ref{integb}) 
can be identified with the elliptic genus of $K3$ due to the following identities
\begin{eqnarray} \label{identek3}
\frac{1}{\eta^{24}} \left[ E_4^2 E_{4, 1} (\tau, z) - E_6 E_{6, 1} (\tau, z) \right]  = 72 Z_{K3}( \tau, z) = 576 A (\tau, z) , \\ \nonumber
\frac{1}{\eta^{24}} \left[ 
E_4^2 E_{4, 1}^ {\rm{even, odd}} - E_6 E_{6, 1}^{\rm{even, odd}} \right] 
= 72 Z_{K3}^{\rm{even, odd}} = 576 A^{\rm even, odd} . 
\end{eqnarray}
where $Z_{K3}(\tau, z) = 8  A(\tau, z) $ is the elliptic genus of $K3$. 
The integral  in (\ref{integb}) can be performed \cite{Kawai:1995hy} and it  results in 
\begin{equation}\label{integb1}
\Delta_{G'}^{(1)}( T, U, V) - \Delta_{G}^{(1)} ( T, U, V) =  -48 {\log }
\left[ ({\rm det}{\rm Im} \Omega)^{10} |\Phi_{10} (T, U , V) |^2  \right] . 
\end{equation}
where $\Phi_{10}(T, U, V)$  is the unique Siegel modular form of weight $10$ 
under $Sp(2, \mathbb{Z})$ which is 
also known as the Igusa cusp form. 
The observation that the difference in thresholds of the two gauge groups 
results in the Igusa cusp form was made in \cite{Stieberger:1998yi}. 
It is also important to note that the duality symmetry $SO(3, 2)$ present 
classical in  heterotic on $K3\times T^2$ is broken to $Sp(2, \mathbb{Z})$ 
due to this quantum correction. 

The modular form $\Phi_{10}(T, U, V) $ also determines the degeneracies of 
$1/4$ BPS dyons in  heterotic string theories compactified on $T^6$ or equivalently 
type II theories on $K3\times T^2$.  Note that these theories are  
${\cal N}=4$  string vacua while we have evaluated the threshold correction  (\ref{integb1}),
in heterotic compactified on $K3\times T^2$ which has ${\cal N}=2$ supersymmetry. 
It is  also interesting that the difference in 
thresholds is in fact sensitive only the elliptic genus of $K3$.
In the next subsections we will generalize this  property of the gauge thresholds 
to  heterotic compactified on the CHL orbifolds of $K3$

\subsection{Thresholds in the $\mathbb{Z}_2$ orbifold}

Let us first evaluate the threshold integrands without the Wilson line turned on for the
$\mathbb{Z}_2$ orbifold of K3. 
As we have seen in the previous subsection, 
the most suitable  form of the new supersymmetric index for this task 
is the expression in (\ref{fullans2}) in terms of the Eisenstein series. 
Let us write in a compact form using the  lattice sums defined in 
(\ref{lsums}). 
\begin{eqnarray} \label{fullans4}
 {\cal Z}_{new}^{(2)} ( q, \bar q ) &=& - 2 \frac{E_4}{\eta^{24}} 
 \left[ \Gamma^{(0,0)}_{2,2} 2 E_6 +  \Gamma^{(0, 1)}_{2,2} \frac{2}{3}( E_6 + 2 \E_2 (\tau) ) E_4) 
 \right. \\ \nonumber 
 & & \qquad \qquad \left. + \Gamma^{(1, 0)}_{2,2} \frac{2}{3} ( E_6 - \E_2(\tfrac{\tau}{2} ) E_4 )
 + \Gamma^{(1, 1) }_{2,2} \frac{2}{3} ( E_6 - \E_2 ( \tfrac{\tau +1}{2} ) E_4 ) 
 \right]. 
\end{eqnarray}
As discussed in the earlier subsection, the 
insertion of $Q^2$ in the construction of the integrand ${\cal B}$ in (\ref{defbg}) 
is done by the action of $\alpha_G q\partial_q $ with $\alpha_G = 1/8$  and $\alpha_{G'}  = 1/12$
when the
derivative acts on  the lattice partition function $E_4$   and $E_6$ respectively. 
This ensures modular invariance of the resulting integrand. 
Let us first evaluate the threshold integral for the gauge group $E_8$. 
For this,  $\alpha_G q\partial_q $  acts only on the first $E_4$ in (\ref{fullans4}).
This results in 
\begin{eqnarray}
 {\cal B}_{E_8}^{(2)} ( q, \bar q ) 
 &=&  -\frac{2}{24 \eta^{24} } ( \tilde E_2  E_4 - E_6) 
    \left[ \Gamma^{(0,0)}_{2,2} 2 E_6 +  \Gamma^{(0, 1)}_{2,2} \frac{2}{3}( E_6 + 2 \E_2 (\tau) ) E_4) 
    \right. \\ \nonumber
& & \qquad \left.  + \Gamma^{(1, 0)}_{2,2} \frac{2}{3} ( E_6 - \E_2(\tfrac{\tau}{2} ) E_4 )
 + \Gamma^{(1, 1) }_{2,2} \frac{2}{3} ( E_6 - \E_2 ( \tfrac{\tau +1}{2} ) E_4 ) 
 \right]. 
\end{eqnarray}
where $\tilde E_2$ is given by (\ref{te2}). 
Similarly the gauge threshold  integrand for the $E_7$ gauge group is  given by 
\begin{eqnarray}
{\cal B}_{E_7}^{(2)} ( q, \bar q ) 
&=& -\frac{2 E_4 }{24 \eta^{24} }
 \left[ \Gamma^{(0,0)}_{2,2} 2 ( \hat E_2  E_6  -E_4^2)+ 
  \Gamma^{(0, 1)}_{2,2} \frac{2}{3}(  \hat E_2 E_6  -E_4^2 
 + 2 \E_2 (\tau) ) ( \hat E_2 E_4  - E_6 )  \right.  \nonumber \\
 & &  \qquad\qquad 
 + \Gamma^{(1, 0)}_{2,2} \frac{2}{3} ( \hat E_2 E_6  -E_4^2  - \E_2(\tfrac{\tau}{2} ) (\hat E_2 E_4 - E_6)  )
\nonumber \\ 
 & &\qquad\qquad  \left. 
 + \Gamma^{(1, 1) }_{2,2} \frac{2}{3} ( \hat E_2 E_6 -E_4^2 - \E_2 ( \tfrac{\tau +1}{2} ) (\hat E_2 E_4 - E_6) 
 \right]. 
\end{eqnarray}
Now upon taking the difference in the threshold integrands we obtain 
\begin{equation}
 {\cal B}_{E_7}^{(2)} - {\cal B}_{E_8}^{(2)} = {144 } \left[
 2 \Gamma^{(0, 0)}_{2, 2}  + \frac{2}{3} \Gamma^{(0, 1) }_{2, 2} 
 + \frac{2}{3} \Gamma^{(1,0)}_{2,2} + \frac{2}{3} \Gamma^{(1, 1)}_{2, 2}  \right]. 
\end{equation}
The modular integral with these difference can be performed using the methods
in \cite{justin1}. The difference in the gauge thresholds is given by 
\begin{equation}
 \Delta_{E_7}^{(2)} - \Delta_{E_8}^{(2)} =  - 48\log \left\{ 
 T_2 ^8 U_2 ^8 |\eta(T) \eta( 2T) |^{16}  | \eta( U) \eta( 2 U) | ^{16}  \right\}. 
\end{equation}

\subsubsection*{Wilson line $V\neq 0$}

Let us  turn on the Wilson line with values in the gauge group $E_8$. 
To write down the modification in the new supersymmetric index it is convenient to 
introduce the Lattice sums  with the Wilson lines. 
Let us define
\begin{eqnarray}
 \Gamma_{3, 2}^{(0, 0) {\rm even} }  &=&   \sum_{\substack{m_1, m_2, n_1, n_2 \in \mathbb{Z}, \\ b \in 2 \mathbb{Z} }}
 q^{\frac{p_L^2}{2} } \bar q ^{\frac{p_R^2}{2} } , \qquad \qquad \qquad
 \Gamma_{3, 2}^{(0, 0) {\rm odd} }  =    \sum_{\substack{m_1, m_2, n_1, n_2 \in \mathbb{Z}, \\ b \in 2 \mathbb{Z} +1}}
 q^{\frac{p_L^2}{2} } \bar q ^{\frac{p_R^2}{2} }, \\ \nonumber
 \Gamma_{3, 2}^{(0, 1) {\rm even} }  &=&  \sum_{\substack{m_1, m_2, n_1, n_2 \in \mathbb{Z}, \\ b \in 2 \mathbb{Z} }}
 q^{\frac{p_L^2}{2} } \bar q ^{\frac{p_R^2}{2} } (-1)^{m_1}  , \qquad \quad 
 \Gamma_{3, 2}^{(0, 1) {\rm odd } }  =   \sum_{\substack{m_1, m_2, n_1, n_2 \in \mathbb{Z}, \\ b \in 2 \mathbb{Z} }}
 q^{\frac{p_L^2}{2} } \bar q ^{\frac{p_R^2}{2} } (-1)^{m_1} , \\ \nonumber
 \Gamma_{3, 2}^{(1, 0) {\rm even} }  &=&   
 \sum_{\substack{m_1, m_2, n_2 \in \mathbb{Z}, \\ n_1 \in \mathbb{Z} + \frac{1}{2}, \  b \in 2 \mathbb{Z} }}
 q^{\frac{p_L^2}{2} } \bar q ^{\frac{p_R^2}{2} } , \qquad \qquad \qquad \ \,
 \Gamma_{3, 2}^{(1, 0) {\rm odd } }  =   \sum_{\substack{m_1, m_2,  n_2 \in \mathbb{Z}, \\
 n_1 \in \mathbb{Z} + \frac{1}{2}, \ 
 b \in 2 \mathbb{Z} +1 } }
 q^{\frac{p_L^2}{2} } \bar q ^{\frac{p_R^2}{2} }, \\ \nonumber
 \Gamma_{3, 2}^{(1, 1) {\rm even} }  &=&   
 \sum_{\substack{m_1, m_2,  n_2 \in \mathbb{Z}, \\
 n_1 \in \mathbb{Z} + \frac{1}{2}, \ 
 b \in 2 \mathbb{Z}   } }
 q^{\frac{p_L^2}{2} } \bar q ^{\frac{p_R^2}{2} } (-1)^{m_1} , \qquad  \quad \  \,
 \Gamma_{3, 2}^{(1, 1) {\rm odd } }  =    \sum_{\substack{m_1, m_2,  n_2 \in \mathbb{Z}, \\
  n_1 \in \mathbb{Z} + \frac{1}{2}, \ 
  b \in 2 \mathbb{Z} +1 } }
 q^{\frac{p_L^2}{2} } \bar q ^{\frac{p_R^2}{2} } (-1)^{m_1}, \\ \nonumber
\end{eqnarray}
where $p_R, p_L$ are the lattice momenta with the Wilson line given in  (\ref{defprplv}). 
The new supersymmetric index  with Wilson line in the $E_8$ gauge group is given by 
\begin{eqnarray}
 {\cal Z}_{\rm{new}}^{(2)} &=& 
  - 2 \frac{E_{4, 1} }{\eta^{24}}  \otimes 
 \left[ \Gamma^{(0,0)}_{3,2} 2 E_6 +  \Gamma^{(0, 1)}_{3,2} \frac{2}{3}( E_6 + 2 \E_2 (\tau) ) E_4) 
 \right. \\ \nonumber 
 & & \qquad  \left. + \Gamma^{(1, 0)}_{3,2} \frac{2}{3} ( E_6 - \E_2(\frac{\tau}{2} ) E_4 )
 + \Gamma^{(1, 1) }_{3,2} \frac{2}{3} ( E_6 - \E_2 ( \frac{\tau +1}{2} ) E_4 ) 
 \right]. 
\end{eqnarray}
 Note here the product $\otimes$ refers to the fact that the even/odd part of the  $E_{4, 1}$ multiplies the 
 even/odd part of the various lattice sums as explained in the earlier subsection. 
 The threshold integrand for the gauge group $E_8$  broken down to $G$ is given by 
 \begin{eqnarray}\label{be82}
  {\cal B}_{G}^{(2)} ( q, \bar q ) 
 &=&  -\frac{2}{24 \eta^{24} } ( \tilde E_2  E_{4, 1} - E_{6, 1} ) \otimes  
    \left[ \Gamma^{(0,0)}_{3,2} 2 E_6 +  \Gamma^{(0, 1)}_{3,2} \frac{2}{3}( E_6 + 2 \E_2 (\tau) ) E_4) \ \ \ \ \nn 
    \right. \\  
& & \qquad \left.  + \Gamma^{(1, 0)}_{3,2} \frac{2}{3} ( E_6 - \E_2(\frac{\tau}{2} ) E_4 )
 + \Gamma^{(1, 1) }_{3,2} \frac{2}{3} ( E_6 - \E_2 ( \frac{\tau +1}{2} ) E_4 ) 
 \right]. 
 \end{eqnarray}
 To obtain this note that the insertion of $Q^2$ to obtain the threshold integrand is realized by  $\alpha_G q\partial_q$
 acting on $E_{4, 1}$ with $\alpha_G = \frac{1}{7}$. 
 The threshold integrand for the gauge group $E_7$ is given by 
 \begin{eqnarray} \label{be72}
  {\cal B}_{G'}^{(2)} ( q, \bar q ) 
&=& -\frac{2 E_{4,1} }{24 \eta^{24} } \otimes 
 \left[ \Gamma^{(0,0)}_{3,2} 2 ( \hat E_2  E_6  -E_4^2)+ 
  \Gamma^{(0, 1)}_{3,2} \frac{2}{3}(  \hat E_2 E_6  -E_4^2 
 + 2 \E_2 (\tau) ) ( \hat E_2 E_4  - E_6 )  \right.  \nonumber \\
 & &  \qquad\qquad 
 + \Gamma^{(1, 0)}_{3,2} \frac{2}{3} ( \hat E_2 E_6  -E_4^2  - \E_2(\tfrac{\tau}{2} ) (\hat E_2 E_4 - E_6)  )
\nonumber \\ 
 & &\qquad\qquad  \left. 
 + \Gamma^{(1, 1) }_{3,2} \frac{2}{3} ( \hat E_2 E_6 -E_4^2 - \E_2 ( \tfrac{\tau +1}{2} ) (\hat E_2 E_4 - E_6) 
 \right]. 
 \end{eqnarray}
Taking the difference in the threshold integrands given in (\ref{be82}) and (\ref{be72}) 
we obtain
\begin{align}
  {\cal B}_{G'}^{(2)} -  {\cal B}_{G }^{(2)} =  \frac{1}{12 \eta^{12}}&  \left\{  
   2 \Gamma_{3, 2}^{(0, 0)} \otimes 
    ( E_{4, 1} E_4^2 - E_{6, 1} E_6 )  \right.  \\ \nonumber
    &  \quad
 +   \frac{2}{3} \Gamma_{3, 2}^{(0, 1)}   \otimes \left[  ( E_{4, 1} E_4^2 - E_{6, 1} E_6 ) 
  + 2 \E_2 (\tau) ( E_{4,1} E_6 - E_{6, 1} E_4 )  \right]  \\ \nonumber
  &   \quad
  + \frac{2}{3}  \Gamma_{3, 2}^{(1,0 )} \otimes 
  \left[ ( E_{4, 1} E_4^2 - E_{6, 1} E_6 )  
  - \E_{2} (\tfrac{\tau}{2})  ( E_{4,1} E_6 - E_{6, 1} E_4 )  \right] \\ \nonumber
  & \quad   \left. 
  + \frac{2}{3}  \Gamma_{3, 2}^{(1,1 )} \otimes 
  \left[ ( E_{4, 1} E_4^2 - E_{6, 1} E_6 )  
  - \E_{2} (\tfrac{\tau+1}{2})  ( E_{4,1} E_6 - E_{6, 1} E_4 )  \right] \right\}. 
\end{align}
We now use the identity in (\ref{identek3}) as well as the following identity verified in 
appendix \ref{app:a}
\begin{eqnarray} \label{identechl}
  \frac{1}{\eta^{24}} \left ( E_{4,1}(\tau, z)  E_6 - E_{6, 1}(\tau, z)  E_ 4  \right)  &=& -144 \frac{\theta_1 (\tau, z) ^2}{\eta^6} 
  = -144 B(\tau, z) , \\ \nonumber
  \frac{1}{\eta^{24}} \left ( E_{4,1}^{\rm even, odd}   E_6 - 
  E_{6, 1}^{\rm{even, odd}}   E_ 4  \right)  &=& -144 \frac{(\theta_1^2)\null^{ \rm even, odd} }{\eta^6} 
  = -144 B^{\rm even, odd}.   
\end{eqnarray}
Substituting the identities (\ref{identek3}) and (\ref{identechl}) we  obtain
\begin{eqnarray}
 {\cal B}_{G'}^{(2)} -  {\cal B}_{G }^{(2)} &=& {24 } \left\{ 
    \Gamma_{3, 2}^{(0, 0)} \otimes 
    4A  +    \Gamma_{3, 2}^{(0, 1)}   \otimes \left[  \frac{4}{3} A - \frac{2}{3} B \E_2 (\tau)  \right]
     \right. \\ \nonumber
  & &  \left. 
  +  \Gamma_{3, 2}^{(1,0 )} \otimes 
  \left[  \frac{4}{3} A + \frac{1}{3} B \E_2 ( \tfrac{\tau}{2} )   \right]    +  \Gamma_{3, 2}^{(1,1 )} \otimes 
  \left[  \frac{4}{3} A + \frac{1}{3} B \E_{2} ( \tfrac{\tau +1}{2} )  \right] \right\}. 
\end{eqnarray}
On comparing the twisted elliptic genus for the $N=2$ CHL orbifold of $K3$ given in 
(\ref{eillipchl2}) we can rewrite the above equation as 
\begin{eqnarray}
{\cal B}_{G'}^{(2)} -  {\cal B}_{G }^{(2)}  &=& {24 }  \left\{ 
    \Gamma_{3, 2}^{(0, 0)} \otimes  F^{(0,0)}  + 
       \Gamma_{3, 2}^{(0, 1)}   \otimes F^{(0,1)}
     +  \Gamma_{3, 2}^{(1,0 )} \otimes  F^{(1,0)}
    +  \Gamma_{3, 2}^{(1,1 )} \otimes F^{(1,1)}
   \right\}. \nonumber \\
\end{eqnarray}
This is precisely the integrand in the modular integral to obtain the Siegel modular
form $\Phi_6(\Omega)$ of weight 6.
Using the result of the integration in \cite{justin1},
we obtain
\begin{equation}\label{siegn2}
\Delta_{G'}^{(2)} ( U, T,  V) - \Delta_{G}^{(2)} ( U, T,  V) = - 48 \log \left[ ( {\rm det\, Im }\Omega)^6 |
\Phi_6( U, T, V) |^2 \right]. 
\end{equation}

 The Siegel modular form, $\Phi_6(T, U, V)$, transforms as a weight $6$ form 
 under a subgroup of $Sp(2, \mathbb{Z}) $. This subgroup is explicitly discussed
 in \cite{justin1}\footnote{See  below equation (3.20) of \cite{justin1}.}. 
 The appearance of the $\Phi_6$  in the threshold calculation here shows
 that the duality group of this compactification is a subgroup of $Sp(2, \mathbb{Z})$. 
 Just as in the case of heterotic string on $K3\times T^2$, 
 the modular form $\Phi_6$ is also related to the partition function 
 of $1/4$ BPS dyons in on type II theory on the CHL orbifold of $K3$. 
 This theory has ${\cal N}=4$ supersymmetry, it is dual to the original CHL compactifications
 of heterotic studied in \cite{Chaudhuri:1995fk}. 
 Let $\tilde \Phi_6$ be the generating function of dyons in 
 this theory, then the modular form $\Phi_6$ is related to $\tilde \Phi_6$  in (\ref{siegn2}) by 
 the following $Sp(2, \mathbb{Z})$ transformation. 
 \begin{equation}
  \Phi_6 (U, T, V)  = T^{-6} \tilde \Phi_6 ( U - \frac{V^2}{T} , - \frac{1}{T}, \frac{V}{T} ) .  
 \end{equation}

\subsection{Thresholds in the $\mathbb{Z}_N$ orbifold}

In this subsection we generalize the calculation of the gauge one loop thresholds 
to the $\mathbb{Z}_N$ orbifold for $N=3, 5, 7$. 
Since we have discussed the case for $N=2$ in detail we will directly present 
the results the threshold with Wilson line embedded in the unbroken gauge group $E_8$. 
Again to present  the results it is convenient to 
define the following lattice sums. 
\begin{align}
\hspace{-.9cm}\Gamma_{3, 2}^{(0, s) {\rm even} }
&=   \sum_{\substack{m_1, m_2, n_1, n_2 \in \mathbb{Z}, \\  b \in 2 \mathbb{Z} }}
 q^{\frac{p_L^2}{2} } \bar q ^{\frac{p_R^2}{2} }  e^{\frac{2\pi i s m_1}{N} }, \quad \ 
 \Gamma_{3, 2}^{(0, s) {\rm odd} }  =   \sum_{\substack{m_1, m_2, n_1, n_2 \in \mathbb{Z}, \\ b \in 2 \mathbb{Z} +1  }}
 q^{\frac{p_L^2}{2} } \bar q ^{\frac{p_R^2}{2} }e^{\frac{2\pi i s m_1}{N} } , \nn \\  
 \Gamma_{3, 2}^{(r , rk ) {\rm even} }
&=   \sum_{\substack{m_1, m_2,  n_2 \in \mathbb{Z}, \\ n_1 \in  \mathbb{Z} + \frac{r}{N} , \ b \in 2 \mathbb{Z} }}
 q^{\frac{p_L^2}{2} } \bar q ^{\frac{p_R^2}{2} }  e^{\frac{2\pi i r k m_1}{N} }, \quad \,
 \Gamma_{3, 2}^{(r, rk) {\rm odd} }  =   \hspace{-.3cm}\sum_{\substack{m_1, m_2, n_2, \in \mathbb{Z}, \\  n_1 \in \mathbb{Z} + \frac{r}{N}, \ 
 b \in 2 \mathbb{Z} +1  }}
 q^{\frac{p_L^2}{2} } \bar q ^{\frac{p_R^2}{2} }e^{\frac{2\pi i r k  m_1}{N} } .
\end{align}
From the expression for the new supersymmetric index in (\ref{fullans3}), it is easy to generalize 
for the situation with the Wilson line embedded in the $E_8$ gauge group. 
This  is given by 
\begin{eqnarray}
 {\cal Z}_{\rm new}^{(N)} &=&- 2\frac{ E_{4, 1} }{\eta^{24} } \otimes \left\{ 
 \Gamma_{3, 2}^{(0,0)}\frac{4}{N}  E_6 
 + \sum_{s =1}^{N-1}  \Gamma_{3, 2}^{(0,s)}\left[ \frac{4}{N(N+1)} E_ 6  + \frac{4}{N+1} \E_N(\tau) E_4 
 \right]  \right.\nonumber  \\ 
 & &  \;\; \;  \qquad + \left.  \sum_{r = 1, k =0}^{N-1} \Gamma_{3, 2}^{(r, rk)} \left[ 
  \frac{4}{N(N+1)} E_ 6  - \frac{2}{N (N+1)} \E_N(\frac{\tau + k}{N} ) E_4 
 \right] \right\}. \nonumber \\
\end{eqnarray}
Again, using the same manipulations to evaluate the difference in the threshold integrands  
for the two gauge groups, we obtain 
\begin{eqnarray}
 {\cal B}_{G'}^{(N)} - {\cal B}_{G}^{(N)}
  &=& 24 \left\{ 
  \Gamma^{(0, 0)}_{3, 2}  \otimes  \frac{8}{N} A 
  + \sum_{s = 1}^{N-1} \Gamma^{(0,s)}_{3, 2} \otimes \left[
  \frac{ 8}{N(N+1)} A - \frac{2}{N(N+1) } \E_N(\tau) B \right]  \right. \nonumber  \\ 
 & & \quad  \left. + \sum_{r=1, k =0}^{N-1} \Gamma^{(r,rk)}_{3, 2} \otimes \left[
  \frac{8}{N(N+1)} A + \frac{2}{N(N+1) } \E_N( \frac{\tau +k}{N} )  B \right] \right\}. \nonumber\\
\end{eqnarray}
Now using the expressions for the twisted elliptic genus for the CHL orbifold of $K3$ 
given in (\ref{ntwist}) we can recast the above expression as 
\begin{eqnarray}
 {\cal B}_{G'}^{(N)} - {\cal B}_{G}^{(N)}
  = 24 \sum_{r, s =0}^{N-1} \Gamma^{(r, s} \otimes  F^{(r, s)}.  
\end{eqnarray}
The integral of this function over the fundamental domain  
has been performed in  \cite{justin1}. 
The result of this integral is 
\begin{equation}
 \Delta_{G'}^{(N)} ( U, T, V) - \Delta_{G}^{(N)} ( U, T, V) 
 = - 48 \log [ ( {\rm det\, Im } \Omega)^k |\Phi_k( U, T, V) |^2 ]  . 
\end{equation}
Here $\Phi_k$ is the Siegel modular form of weight $k$ transforming according to a subgroup of $Sp(2, \mathbb{Z})$. 
This modular form is related to $\tilde \Phi_k$ the generating function for 
$1/4$ BPS dyons in type II theory compactified on the CHL orbifold of $K3$  by the $Sp(2, \mathbb{Z})$ transformation
\begin{equation}
   \Phi_k (U, T, V)  = T^{-k} \tilde \Phi_k ( U - \frac{V^2}{T} , - \frac{1}{T}, \frac{V}{T} ) .  
\end{equation}

We have thus demonstrated that the moduli dependence in the  difference in the gauge thresholds for 
heterotic string compactified on the CHL orbifold of $K3$ are captured by 
Siegel modular forms $\Phi_k$ of weight $k = \frac{24}{(N+1)} - 2$. These are related to the modular forms 
which are generating functions for $1/4$ BPS states in ${\cal N}=4$ string  theories 
obtained by compactifying type II theories on the CHL orbifold of $K3$.

\section{Conclusions}\label{sec:6}

We have introduced ${\cal N}=2$ string theories constructed by compactifying heterotic 
string theories on CHL orbifolds of $K3$ .
These generalize the well studied example of the heterotic string  compactified 
on $K3\times T^2$. 
The  CHL orbifolding reduces the number of hypers in the  resulting ${\cal N}=2$ theory and 
preserves the vectors in the theory. 
These models do not have a lift to $6$ dimensions since the orbifolding involves a
shift on one of the circles of $T^2$. 
We evaluated the new supersymmetric index for these compactifications 
and showed that it admits an expansion in terms of the McKay-Thompson 
series  of the group $M_{24}$ associated with the $\mathbb{Z}_N$ involution used to 
construct the CHL orbifold. 

We then studied the moduli dependence of one-loop corrections to the gauge couplings 
in the CHL orbifolds of $K3$. We showed that the moduli dependence 
of the difference in the gauge thresholds is captured by Siegel modular  forms 
closely related to partition function of $1/4$ BPS dyons in ${\cal N}=4$ string theories.
These Siegel modular forms transform under sub-groups of $Sp(2, \mathbb{Z})$ 
which shows that the the CHL orfbifolding reduces the duality 
symmetry of the original $K3$ compactification to a subgroup of $Sp(2, \mathbb{Z})$.

 It will be interesting to evaluate gravitational thresholds in these theories 
 to see if these also admit a nice  structure seen for the gauge thresholds. 
Another direction to explore is generalize the observations of this paper to other examples. 
A simple example  
to study is the compactification in the heterotic string 
which will lead to the Siegel modular which captured degeneracies of 
dyons in type II ${\cal N}=4$ constructed in \cite{justin2}. 
Another generalization is to consider compactifications in heterotic based 
on the  new classes of  twisted elliptic genera of $K3$ constructed in \cite{Cheng:2010pq,Gaberdiel:2010ch,Eguchi:2010fg}.

We  observed  that the difference in integrands of the gauge thresholds
 reduces to the twisted elliptic genus of $K3$ for the CHL orbifold.
 This points to the fact that the 
difference in the  thresholds is essentially sensitive only to  a supersymmetric index of the 
  internal CFT. 
It will be interesting to prove this in general. 
A similar phenomenon was observed by \cite{Angelantonj:2014dia,Angelantonj:2015nfa}, in which the authors 
evaluated the difference in thresholds in compactifications of heterotic 
which completely break supersymmetry.
They noticed  that the difference in thresholds is purely a holomorphic function
in the modular parameter indicative of a supersymmetric index. 

Another direction worth exploring is the ${\cal N}=2$ string duality between 
heterotic string theory compactified on these CHL orbifolds of $K3$ and 
the appropriate Calabi-Yau on the type II side. 
Since the CHL orbifolds reduce the number of hypers, the appropriate Calabi-Yau
should have the reduced Hodge number $h_{2,1} = 6k + 4 $. It is interesting to study 
what symmetry action on the Calabi-Yau  reproduces this Hodge number. 
In this context it will be also important to study the one-loop threshold corrections to gravitational
couplings in these models. 
Note that the modular forms $\Phi_k$ obtained in the difference of thresholds 
of the CHL compactifications in this paper factorize in the $V\rightarrow 0$ limit 
as \cite{justin1} 
\begin{equation}
 \lim_{V\rightarrow 0} \Phi_k( U, T, V) \sim V^2 ( \eta( T) \eta(N T) )^{k+2}   ( \eta( U) \eta(N U) )^{k+2} . 
\end{equation}
It is also interesting to investigate if the difference in thresholds
have other  degeneration limits  for discrete values of $V$ as seen in 
\cite{Angelantonj:2014dia,Angelantonj:2015nfa} \footnote{We thank 
Ioannis Florakis for raising this point.}. 
This degeneration should correspond to charged states becoming massless
since it corresponds to a logarithmic singularity in the one-loop threshold. 
It 
will be interesting to explore this phenomenon on the dual Calabi-Yau compactification
in type II.  

\acknowledgments

J.R.D thanks Edi Gava, Dileep Jatkar, K.S. Narain and Ashoke Sen for  discussions related to this problem at 
several instances over the last few years. 
He  thanks Shailesh Lal and Dileep Jatkar for initially  collaborating  on this project. 
He also thanks the hospitality of the string group at MPI, M\"{u}nchen during which 
this collaboration was initiated. 
We are very grateful to Ioannis Florakis for a careful reading of the manuscript
and suggestions which helped to improve it. 
Furthermore we  thank Stephan Stieberger for discussions,  encouragement to pursue  and complete this project and 
also for reading the manuscript. 
A major part of this work was completed when S.D was supported by a research associateship of the Indian Institute of Science. 
The research of S.D is presently supported by the NCCR SwissMAP, funded by the Swiss
National Science Foundation. The work of D.L.  is partially supported by the ERC Advanced Grant "Strings and Gravity"
(Grant.No. 32004) and by the DFG cluster of excellence "Origin and Structure of the Universe".

\appendix
\def\bhalf{\frac{1}{2}}

\section{Theta functions and  Einsenstein Series}  \label{app:a}

Our notations for the 
the Jacobi theta functions are summarized in the following  expansion
\begin{eqnarray}
\theta_1 (  \tau, z) &=&  \sum_{n\in \mathbb{Z}} \exp
\left[ i \pi ( n + \tfrac{1}{2} )^2 \tau + 2\pi i ( n + \tfrac{1}{2} ) ( z+ \tfrac{1}{2} ) \right] , \\ \nonumber
\theta_2 (  \tau, z ) &=&  \sum_{n\in \mathbb{Z}} \exp
\left[ i \pi ( n + \tfrac{1}{2} )^2 \tau + 2\pi i ( n + \tfrac{1}{2} ) z \right] , \\ \nonumber
\theta_3 ( \tau, z ) &=&  \sum_{n\in \mathbb{Z}} \exp
\left[ i \pi   n ^2 \tau + 2\pi i  n  z  \right] , \\ \nonumber
\theta_4 (  \tau, z) &=&  \sum_{n\in \mathbb{Z}} \exp
\left[ i \pi  n ^2 \tau + 2\pi i  n  ( z+ \tfrac{1}{2} ) \right] . \\ \nonumber
\end{eqnarray}
When there is no ambiguity we will use the following notation for  theta functions at the origin
\begin{eqnarray}
 \theta_2(\tau, 0 ) = \theta_2(q) = \theta_2, \quad \theta_3(\tau, 0) = \theta_3(q) = \theta_3, \quad
 \theta_4(\tau, 0) = \theta_4(q) = \theta_4. 
\end{eqnarray}
The Dedekind $\eta$ function is defined by the product
\begin{equation}
 \eta(\tau) = q^{\frac{1}{24}} \prod_{n=1}^\infty ( 1 - q^{n}) , 
\end{equation}
where $q = e^{2\pi i \tau}$. 
It is also useful  to present the infinite product representation of the theta functions at the origin
\begin{eqnarray}
 \frac{\theta_2 (\tau) }{\eta(\tau) } &=&  q^{\frac{1}{12} } \prod_{n=1}^\infty ( 1+ q^n) ( 1+ q^{n-1}) , 
 \\ \nonumber
 \frac{\theta_3 (\tau) }{\eta(\tau) }&=&  q^{-\frac{1}{24} } \prod_{n=1}^\infty ( 1+ q^{n- \frac{1}{2} } ) ( 1+ q^{n-\frac{1}{2} }) , 
 \\ \nonumber
 \frac{\theta_2 (\tau) }{\eta(\tau) }&=&  q^{-\frac{1}{24} } \prod_{n=1}^\infty( 1- q^{n- \frac{1}{2} } ) ( 1- q^{n-\frac{1}{2} }) . 
\end{eqnarray}
One identity of theta functions which we repeatedly use is triple product identity 
\begin{equation}\label{tripprod}
 \theta_2 \theta_3 \theta_4 = 2 \eta^3. 
\end{equation}
Finally we will also use the following shift properties of the theta functions.
\begin{align}
 &\theta_4( \tau, z+ \tfrac{1}{2}) = \theta_3(\tau, z ), \  \qquad\qquad\quad\  \ \ \theta_1( \tau, z+ \tfrac{\pi}{2} ) = \theta_2(\tau, z ), \\ \nonumber
 &\theta_2(\tau, z+ \tfrac{1}{2} ) = -\theta_1(\tau, z) ,   \; \quad \  \qquad\qquad \theta_3( \tau, z + \tfrac{1}{2}) = \theta_4(\tau, z) , \\ \nonumber
 &\theta_4(\tau, z + \tfrac{\tau}{2} ) = i e^{ - \frac{ \pi i \tau}{4}  - i \pi z } \theta_1 (\tau, z) , 
   \qquad \, 
 \theta_1 (\tau, z + \tfrac{\tau}{2} ) = i e^{ - \frac{ \pi i \tau}{4}  - i \pi z } \theta_4 (\tau, z), \\ \nonumber
 &\theta_2  (\tau, z + \tfrac{\tau}{2} ) =  e^{ - \frac{ \pi i \tau}{4}  - i \pi z } \theta_3 (\tau, z), 
   \qquad\ \,
 \theta_3 (\tau, z + \tfrac{\tau}{2} ) =  e^{ - \frac{ \pi i \tau}{4}  - i \pi z } \theta_4 (\tau, z). 
\end{align}

The Eisenstein series $E_2$ is a quasi-modular form whose series expansion is given by 
\begin{equation}
 E_2 (q) = 1 - 24 \sum_{n=1}^\infty \sigma_1(n) q^n. 
\end{equation}
where $\sigma_1(n)$ is the sum of positive integral divisors of $n$. 
The combination 
\begin{equation}
 \tilde E_2   = E_2 - \frac{3}{\pi \tau_2},  
\end{equation}
transforms as a good modular form of weight $2$. 
The Einsenstein series $E_4$ and $E_6$ are related to the theta functions by the well known identities
\begin{eqnarray}
 E_4 &=&  \frac{1}{2} \left( \theta_2^8 + \theta_3^8 + \theta_4^8 \right), \\ \nonumber
 E_6 &=&  \frac{1}{2} \left[ - \theta_2^6( \theta_3^4 + \theta_4^4) \theta_2^2 + \theta_3^6( \theta_4^4 - \theta_2^4) \theta_3^2
 + \theta_4^6( \theta_2^4 + \theta_3^4) \theta _4^2 \right] . 
\end{eqnarray}
We will also require the modular form 
$\E_N$ by 
\begin{eqnarray}
\E_N (\tau) = \frac{12 i}{\pi (N-1)} \pd_\tau \log \frac{\eta(\tau)}{\eta(N\tau)}. 
\end{eqnarray}
Under modular transformations it behaves as
\begin{equation}
 \E_N(\tau + 1) = \E_N , \qquad \E_N( -1/\tau) = - \tau^2\frac{1}{N}  \E_N( {\tau}/{N} ) . 
\end{equation}

The  relations given in (\ref{mident}) involving   $\theta$  functions Eisenstein series and the $\E$ function 
is used to write the new supersymmetric in terms of the Eisenstein series. 
We have established these identities by performing $q$ expansions 
in Mathematica, we have listed out the first few terms
\begin{align}
-(\th_3^8 \th_4^4 + \th_4^8 \th_3^4 ) &= -\frac{2}{3} \left(E_6 + 2 \E_2(\tau) E_4 \right) \\ \nn 
&=2+16 q-496 q^2+3904 q^3-16880 q^4+50016 q^5-121024 q^6+\cdots,  
\end{align}
\begin{align}
\th_3^8 \th_2^4 + \th_2^8 \th_3^4  &= -\frac{2}{3} \left(E_6 - \E_2\left(\tfrac{\tau}{2}\right) E_4 \right) \\
&= 16 q^{1/2}+512 q+3904 q^{3/2}+16384 q^2+50016 q^{5/2}+124928 q^3+\cdots,  \nn 
\end{align}
\begin{align}
\th_2^8 \th_4^4 - \th_2^8 \th_4^4  &= -\frac{2}{3} \left(E_6 - \E_2\left(\tfrac{\tau+1}{2}\right) E_4 \right) \\
&=   16  q^{1/2}-512 q+3904 q^{3/2}-16384 q^2+50016 q^{5/2}-124928 q^3 	+ \cdots.  		\nn 
\end{align}

Finally we establish the identities in (\ref{identek3}) and  (\ref{identechl}). 
First recall that the Jacobi forms of index 1 admit a even odd  decomposition 
given by 
\begin{align}
f(\tau, z) =  f^{\rm even}  (\tau ) \theta_{\rm even}   +  f^{\rm odd}  (\tau ) \theta_{\rm odd} , 
\end{align}
where 
\begin{equation}
\label{expeo}
f^{\rm even}  = \sum_{N\equiv 0(4)} c(N) q^{N/4} \qquad f^{\rm odd} = \sum_{N\equiv -1(4)} c(N) q^{N/4},  
\end{equation}
and 
\begin{equation}
 \theta_{\rm even}(\tau, z) = \theta_3 (2\tau, 2 z), \qquad 
 \theta_{\rm odd}(\tau, z) = \theta_2 (2\tau, 2 z). 
\end{equation}
It is clear that $E_{4, 1}( \tau, z)$ and $E_{6, 1}(\tau, z)$ defined in (\ref{defe41}) and ({\ref{defe61})  admit 
this  decomposition using the identities given in (\ref{thetasq}). 
From these we find that 
\begin{eqnarray}
E_{4,1}^{\rm even}  &=& \theta_3^7(2\tau ) + 7 \theta_3^3 (2\tau ) \theta_2^4 (2\tau), \nn \\
E_{4,1}^{\rm odd}  &=&  \theta_2^7(2\tau ) + 7 \theta_2^3 (2\tau ) \theta_3^4 (2\tau). 
\end{eqnarray}
while for $E_{6,1}$ it is
\begin{align}
E_{6,1}^{\rm even} 
&= 	\th_3^{11}(2\tau) - 11 \, \th_2^8(2\tau) \th_3^3(2\tau) - 22\,  \th_2^4(2\tau) \th_3^7(2\tau), 		\nn \\
E_{6,1}^{\rm odd}  &= \th_2^{11}(2\tau) - 11 \, \th_3^8(2\tau) \th_2^3(2\tau) - 22\,  \th_3^4(2\tau) \th_2^7(2\tau). 	
\end{align}
From (\ref{expeo}) we see that  $q$ expansions of 
 the `even'  and  `odd' parts of the Jacobi forms are different
 therefore we can introduce the notation \cite{Cardoso-Curio-Lust} 
 in which we combine these expansions 
 \begin{align}
 \widehat{f}(\tau) = f^{\rm even} (\tau) + f^{\rm odd}  (\tau ). 
 \end{align}
Then we establish (\ref{identek3}}) and (\ref{identechl} by performing the $q$ expansions in Mathematica 
which are given by 
\begin{align}
& 8 \left( \frac{\widehat\theta_2^2 }{\theta_2^2 }  + \frac{\widehat\theta_3^2 }{\theta_3^2 }+
\frac{\widehat\theta_4^2 }{\theta_4^2 }  \right)  = \frac{1}{72} \, 
\frac{E_4^2 \widehat{E_{4,1}} - E_6 \widehat{E_{6,1}}}{\eta^{24}}  \\
& = {2\over q^{1/4} }+ 20 - 128 q^{3/4} + 216 q - 1026 q^{7/4} + 1616 q^2 - 
5504 q^{11/4} + 8032 q^3	+ \cdots , 		\nn 
\end{align}
and 
\begin{align}
{-2}\, \frac{\widehat\theta_1^2 }{\eta^6}  &= \frac{1}{72} \, \frac{E_6 \widehat{E_{4,1}} - E_4 \widehat{E_{6,1}}}{\eta^{24}}  \\
&={2\over q^{1/4}}-4+16 q^{3/4}-24 q+78 q^{7/4}-112 q^2+304 q^{11/4}-416 q^3 + \cdots. 	\nn
\end{align}

\section{Lattice sums}\label{app:b}

In this appendix we provide the details of evaluating the lattice sum over the shifted lattice $E_8'$ defined by 
\begin{equation} \label{defp}
 P_{(a, b)} = e^{-2\pi i \frac{ab}{n^2 } \gamma^2} \sum_{\lambda \in \Gamma^8  + \frac{a}{2} \gamma}
 e^{ 2\pi i \frac{b}{n} \lambda \cdot \gamma} q^{\frac{1}{2} \lambda^2}. 
\end{equation}
The sum runs over all the lattice vectors $\lambda$ of $E_8$. 
The lattice shift $\gamma$ of the $\mathbb{Z}_2$ orbifold is given by 
\begin{equation} \label{defgam}
 \gamma = ( 1, 1,  0^6) . 
\end{equation}
Before we proceed let us recall that the roots  of $E_8$ are given by 
\begin{align}
&\text{112 root vectors of } D_8 : (\dots,\pm 1, \dots , \pm 1, \dots ),  \nn \\
&\text{128 8-dimensional vectors } : \left( \pm \half,  \pm \half ,  \pm \half , \dots  \right).  \nn 
\end{align}
Here the `$\dots$' in the 112 root vectors of $D_8$ represent zeros. 
The lattice  vectors are then of two types. 
\begin{align} \label{wt-vectors}
\lambda_A &= (n_1, n_2, \cdots, n_8),  \nn \\
\lambda_B &= \left(  n_1 +  \tfrac{1}{2}, \cdots , n_8 +  \tfrac{1}{2} \right), 
\end{align}
with the constraint 
\begin{align}\label{n-constraint}
\sum_{i=1}^8 n_i = \text{even integer} . 
\end{align}

\def\Z{\mathbb{Z}}
\def\vth{\vartheta}

Let us now perform the lattice sum without any shifts. This is the $(0,0)$  sector. 
\begin{align}
P_{(0, 0) } = \sum_{\lambda_A} q ^{\half \lambda_A \cdot \lambda_A}  +  \sum_{\lambda_B} q ^{\half \lambda_B \cdot \lambda_B} . 
\end{align}
We can impose the constraint via an extra factor 
\begin{align*}
\half (1+ e^{i\pi \sum n_i}) = \begin{cases}
1  \quad \text{if $\sum_i n_i$ = even integer } \\
0  \quad \text{otherwise}. 
\end{cases}
\end{align*}
This results in 
\begin{align}
P_{(0, 0) } \ = \ \  &\bhalf \prod_{i=1}^8 \sum_{n_i \in \Z } e^{i\pi \tau n_i^2 } + 
\bhalf \prod_{i=1}^8 \sum_{n_i \in \Z } e^{i\pi n_i^2 }  e^{i\pi \tau n_i } \nn \\ &+
\bhalf\prod_{i=1}^8 \sum_{n_i \in \Z } e^{i\pi \tau (n_i+\half)^2 } +
\bhalf\prod_{i=1}^8 \sum_{n_i \in \Z } e^{i\pi \tau (n_i+\half)^2 }e^{i\pi  (n_i+\half)}, 
\end{align}
which can be written in terms of the Jacobi $\theta$-functions as 
\begin{align}
P_{(0, 0) } = \bhalf \left[ \theta_3^8    +  \theta _4^8    +    \theta_2^8   + \theta_1^8   \right]. 
\end{align}
The last term is zero. Hence the final expression for the lattice sum is
\begin{align}
P_{(0, 0) } = \bhalf \left[\theta_2^8+\theta_3^8+\theta_4^8\right]. 
\end{align}

For  the case $(a,b)=(0,1)$.  The weight vectors are the same as \eqref{wt-vectors}.
To evaluate the phase we 
use the shift  in (\ref{defgam}) and the weight vectors  to get 
\begin{equation}
\lambda_A  \cdot \gamma = 
n_1 +n_2  , \qquad 
\lambda_B  \cdot \gamma = 
(n_1 + \tfrac{1}{2})+ (n_2 +  \tfrac{1}{2}). 
\end{equation}
The lattice sum is 
\begin{align}
P_{(0, 1) } =
\ &\bhalf \prod_{i=1}^{6} \sum_{n_i \in \Z} e^{\pi i \tau n_i^2 }   \, \prod_{i=1}^{2}
\sum_{n_i \in \Z} e^{\pi i \tau n_i^2 } e^{\pi i  n_i } +  \bhalf \prod_{i=1}^{6} 
\sum_{n_i \in \Z} e^{\pi i \tau n_i^2 }   \, \prod_{i=1}^{2} \sum_{n_i \in \Z} e^{\pi i \tau n_i^2 } e^{2\pi i n_i }  \nn \\
&+ \bhalf \prod_{i=1}^{6} \sum_{n_i \in \Z} e^{\pi i \tau (n_i+\half )^2 }   \, 
\prod_{i=1}^{2} \sum_{n_i \in \Z} e^{\pi i \tau  (n_i+\half )^2 } e^{\pi i   (n_i+\half )} \nn \\ &+  
\bhalf \prod_{i=1}^{6} \sum_{ n_i\in \Z} e^{\pi i \tau (n_i+\half )^2 }   \, \prod_{i=1}^{2}
\sum_{n_i \in \Z} e^{\pi i \tau  (n_i+\half )^2 } e^{2\pi i (n_i+\half ) },  \nn \\ 
= \ & \bhalf \left[	\theta_3^6 \theta_4^2 + \theta_4^6 \theta_3^2 + \theta_2^6 \theta_1^2 -\theta_2^2 \theta_1^6	\right] \ 
= \ \bhalf \left[	\theta_3^6 \theta_4^2 + \theta_4^6 \theta_3^2 	\right]. 
\end{align}

For $(a,b)=(1,0)$, the weight vectors are
\begin{align}\label{wt-2}
\lambda_A' &= (n_1 + \half, n_2+ \half, n_3, n_4 \cdots, n_8),  \nn \\
\lambda_B '&= \left(  n_1 + 1 , n_2 +1, n_3+\half , \cdots , n_8 + \half \right). 
\end{align}
The lattice sum is then
\begin{align}
P_{(1, 0) } = \ & \bhalf\prod_{i=1}^{2} \sum_{n_i \in \Z} e^{i \pi \tau (n_i+\half)^2 	}
\prod_{i=1}^{6} \sum_{n_i \in \Z} e^{i \pi \tau n_i^2 	} + 
\bhalf \prod_{i=1}^{2} \sum_{n_i \in \Z} e^{i \pi \tau (n_i+\half)^2 	}  e^{i \pi n_i  }  
\ \prod_{i=1}^{6} \sum_{n_i \in \Z} e^{i \pi \tau n_i^2 	} e^{i \pi n_i}\nn \\
&  + \bhalf\prod_{i=1}^{2} \sum_{n_i \in \Z} e^{i \pi \tau n_i^2 	}  \prod_{i=1}^{6} 
\sum_{n_i \in \Z} e^{i \pi \tau (n_i+\half)^2 	} +
\bhalf   \prod_{i=1}^{2} \sum_{n_i \in \Z} e^{i \pi \tau n_i^2 	}   e^{i \pi n_i	}\prod_{i=1}^{6} 
\sum_{n_i \in \Z} e^{i \pi \tau (n_i+\half)^2 	} e^{i \pi n_i	},  \nn \\
= \ & \bhalf \left[   \theta_3^6 \theta_2^2 + \theta_2^6 \theta_3^2 - \theta_4^6 \theta_1^2 - \theta_1^6 \theta_4^2  		\right]
= \bhalf\left[  \theta_3^6 \theta_2^2 + \theta_2^6 \theta_3^2 \right] . 
\end{align}

Finally for $(a,b)=(1,1)$ the weight vectors are same as the ones in equation \eqref{wt-2}. 
In addition we also have the extra phase since $b\neq 0$. Here
\begin{equation}
\lambda_{A}' \cdot \gamma =
(n_1 + \half ) + (n_2 + \half ) \qquad 
\lambda_{B}' \cdot \gamma = 
n_1 + n_2 
\end{equation}
So the lattice sum in this case is  given by 
\begin{align}
- P_{(1, 1) } = & \ \bhalf \prod_{i=1}^2 \sum_{n_i \in \Z} e^{\pi i \tau (n_i + \half )^2 + \pi i (n_i +\half )} \prod_{i=1}^6 \sum_{n_i \in \Z} e^{\pi i n_i^2 } \nn \\ 
& + \bhalf \prod_{i=1}^2 \sum_{n_i \in \Z} e^{\pi i \tau (n_i + \half )^2 + \pi i n_i + \pi i (n_i + \half)} \prod_{i=1}^6 \sum_{n_i \in \Z} e^{\pi i n_i^2 + \pi i n_i } \nn \\
& + \bhalf  \prod_{i=1}^2 \sum_{n_i \in \Z} e^{\pi i \tau n_i^2} e^{\pi i n_i }  \prod_{i=1}^6 \sum_{n_i \in \Z}	e^{\pi i \tau (n_i + \half )^2  }  					\nn \\ 
& + \bhalf \prod_{i=1}^2  \sum_{n_i \in \Z} e^{\pi i \tau n_i^2 + \pi i n_i + 
\pi i n_i }\prod_{i=1}^6 \sum_{n_i \in \Z} e^{\pi i \tau (n_i +\half )^2  +  \pi i (n_i + \half)}, \nn  \\
= &  \ \bhalf \left[ \theta_3^6 \theta_1^2 - \theta_4^6 \theta_2^2 + \theta _2 ^6 \theta_4 ^2    - \theta_1^6 \theta_3 ^2 \right] \ 
= \  \bhalf \left[ \theta _2 ^6 \theta_4 ^2 - \theta_4^6 \theta_2^2  \right] . 
\end{align}
Note that is the case where there are corrections due to the shift 
and factors present due to the even integer constraint and the extra phase. 
The overall negative sign is due to the overall phase in the definition (\ref{defp}).

\section{Details for the $\mathbb{Z}_2$ orbifold} \label{app:c}

\def\cF{\mathcal{F}}
\def\mmnn{{m_1,m_2,n_1,n_2}}

This appendix provides the details of the evaluation of the following trace
\begin{equation}
 F_{m_1, m_2, n_1, n_2} ( a, r, b, s;q) = {\rm Tr}_{m_1, m_2, n_1, n_2; g^a g^{\prime r}; RR} 
 \left(  g^b g^{\prime  s}  e^{i\pi ( F^{T^4}  +  F^{T^2}) } F^{T^2}  q^{L_0'} \bar q^{\bar L_0'} 
 \right) . 
\end{equation}
The orbifold action $g$ and $g'$ is defined in (\ref{orbiact}).  
We label the various sectors in terms of only the action of $g$. 
The action of $g'$ is summed over in each of these sectors. 
Also in the above trace the bosonic oscillators in the holomorphic direction of the $T^2$ 
is not included since it is has already been included in (\ref{indchl}). 
Due to the presence of the fermionic zero modes associated 
with  $T^4$  which is along the $6, 7, 8, 9$ direction
the trace vanishes for 
$a=0, b=0$ irrespective of the values of $r$ and $s$. 
Therefore we have 
\begin{equation}
  F_{m_1, m_2, n_1, n_2} ( 0, r, 0, s;q) =0. 
\end{equation}
Therefore from the definition of ${\cal F}$ in (\ref{tracdef}) 
we see that  this implies 
\begin{equation}
 {\cal F}_{m_1, m_2, n_1, n_2} (0, 0; q) =0. 
\end{equation}

Now let us move to the $(0, 1)$ sector. 
We have the following
\begin{eqnarray}\label{f1}
  F_{m_1, m_2, n_1, n_2} ( 0, 0, 1, 0; q)  &=&- 4 \frac{1}{ [  q^{\frac{1}{24}}  \prod_{n=1}^\infty ( 1+ q^n)  ]^4 } 
  \\ \nonumber& = & - 4 \frac{\theta_3^2 \theta_4^2}{\eta^4} . 
\end{eqnarray}
In the above equation the factor
 $4$ arises from the anti-holomorphic fermion zero modes associated with the $T^4$. 
The bosonic and the fermionic oscillators in the anti-holomorphic sector cancel. 
What is left behind  are the  $4$  bosonic oscillators in the holomorphic sector. The action of 
$g$ on these oscillators reverses the sign.  We have used the  product representation of the $\theta_2$
and then  triple product identity in (\ref{tripprod}) to arrive at (\ref{f1}). 
The over all negative sign is associated with the action of$g$  
 on the vacuum. This choice of the action of $g$  on the vacuum 
ensures the final result is modular invariant. 
Next we have
\begin{eqnarray}
 F_{m_1, m_2, n_1, n_2} ( 0, 0, 1, 1; q)  & =& -  (-1)^{m_1}  4 \frac{\theta_3^2 \theta_4^2}{\eta^4} . 
\end{eqnarray}
The only difference in  this trace from that of (\ref{f1}) is the insertion of $g'$  in the trace. 
This picks up the factor $(-1)^{m_1}$ on the state carrying $m_1$ units of momentum 
along the circle $y^4$. 
Now the following traces vanish
\begin{eqnarray}
 F_{m_1, m_2, n_1, n_2} ( 0, 1, 1, 0; q) = F_{m_1, m_2, n_1, n_2} ( 0, 1, 1, 1; q) = 0.  
\end{eqnarray}
This is because in this sector the winding numbers along $y^6$ is half integer moded and 
the action of $g$ as well as $gg'$ reverses the sign of these modes and therefore they
do not contribute in the trace. 
Thus we have
\begin{eqnarray}
{\cal F}_{m_1, m_2, n_1, n_2} ( 0, 1;q)  & = & \begin{cases}
  - 2 \left(1+(-1)^{m_1}\right ) \frac{\th_3 ^2 \th_4 ^2}{\eta^{4}} \quad   &\text{for } \{m_1, m_2, n_1, n_2\}  \in  \mathbb{Z}, \\
    0  \quad  & \text{for }  \{m_1, m_2, n_2\}  \in  \mathbb{Z}, \;\;  \{n_1\} \in \mathbb{Z } + \frac{1}{2}.
 \end{cases}
\end{eqnarray}

For the twisted $(1, 0)$  sector  we have
\begin{eqnarray}\label{f2}
F_{m_1, m_2, n_1, n_2} ( 1, 0, 0, 0; q ) &=&  16 
\frac{1}{ [q^{-\frac{1}{48} } \prod_{n=1}^\infty ( 1 - q^{n-1/2} )   ]^4 } , \\ \nonumber
&=& 4 \frac{\theta_2^2 \theta_3^2}{\eta^4}. 
\end{eqnarray}
Here the factor of $16$ in the first line is due to the 16 twisted sectors 
localized at the 16 fixed points of $T^4$ at $y^{m} = 0, \pi$ for $m = 6, 7, 8, 9$. 
To arrive at the second line we have used the product representation of 
$\theta_4$ and the identity (\ref{tripprod}). 
Again the bosonic and fermionic oscillators in the anti-holomorphic sector cancel
leaving behind the bosonic oscillators in the holomorphic sector. 
These oscillators are half integer modded since they belong to the twisted sector. 
Now 
\begin{eqnarray}
F_{m_1, m_2, n_1, n_2} ( 1, 0, 0, 1; q )   =0. 
\end{eqnarray}
This is because the action of $g'$ exchanges the fixed points pairwise, 
the twisted sector states are off diagonal and  therefore the trace vanishes. 
\begin{eqnarray}
F_{m_1, m_2, n_1, n_2} ( 1, 1, 0, 0 ; q ) = 4 \frac{\theta_2^2 \theta_3^2}{\eta^4} .
\end{eqnarray}
Here the states twisted by $gg'$ are now labelled by the 
fixed points $y^6 = \frac{\pi}{2} , \frac{3\pi}{2} , y^m = 0, \pi$ for $m = 7, 8, 9 $. 
The rest of the analysis to obtain the above equation is same as that in (\ref{f2}), but
note that here the winding $n_1 \in \mathbb{Z} + \frac{1}{2}$ due to the twisting by $g'$. 
Finally 
\begin{eqnarray}
F_{m_1, m_2, n_1, n_2} ( 1, 1, 0, 1 ; q ) =0. 
\end{eqnarray}
This is because the action of the $g'$ insertion exchanges the fixed points
and the elements are off diagonal in the trace. 
In summary the contributions in this sector are
\begin{eqnarray}
{\cal F}_{m_1, m_2, n_1, n_2} ( 1, 0 ;q) & =&  \begin{cases} 
 2  \frac{\th_2^2 \th_3^2}{\eta^4}   \quad   &\text{for } \{m_1, m_2, n_1, n_2\}  \in  \mathbb{Z},  \\
  2  \frac{\th_2^2 \th_3^2}{\eta^4}  \quad  & \text{for }  \{m_1, m_2, n_2\}  \in  \mathbb{Z},  \{n_1\} \in \mathbb{Z } + \frac{1}{2}.  
\end{cases}
\end{eqnarray}

Lets now look at the $(1, 1)$ sector. 
We have
\begin{eqnarray} \label{f3}
F_{m_1, m_2, n_1, n_2} ( 1, 0, 1, 0 ; q )  &=& - 16 
\frac{1}{ [q^{-\frac{1}{48} } \prod_{n=1}^\infty ( 1 - q^{n-1/2} )   ]^4 }, \\ \nonumber
= - 4 \frac{\theta_2^2 \theta_4^2}{\eta^4} . 
\end{eqnarray}
Again due to the  arguments mentioned earlier, it is only the bosonic oscillators in the $T^4$ directions
which contribute. The $16$ in the first line is due to the presence of the 
$16$ fixed point and the negative sign is because the action of $g$ on the vacuum 
gives a negative sign. 
Now
\begin{equation}
F_{m_1, m_2, n_1, n_2} ( 1, 0, 1, 1 ; q ) = 0. 
\end{equation}
This is because the insertion of $gg'$ in the trace exchanges the 
fixed points pair wise and therefore the elements are off diagonal in the trace. 
Again  due to the same reason of the elements being off diagonal we have
\begin{equation}
F_{m_1, m_2, n_1, n_2} ( 1, 1, 1, 0 ; q ) = 0. 
\end{equation}
Note here the due to twisted by $g'$ the states are at 
$y^6 = \frac{\pi}{2} , \frac{3\pi}{2} , y^m = 0, \pi$ for $m = 7, 8, 9 $. 
Finally 
\begin{equation}
F_{m_1, m_2, n_1, n_2} ( 1, 1, 1, 1 ; q ) = - 4 ( -1)^{m_1}  \frac{\theta_2^2 \theta_4^2}{\eta^4} . 
\end{equation}
Here the analysis is same as in the case of (\ref{f3}). 
The $(-1)^{m_1}$ occurs due to the presence $g'$ in the trace.
Also $n_1 \in \mathbb{Z}+\frac{1}{2}$ since the states are twisted by $g'$. 
To summarize this sector results in 
\begin{eqnarray}
{\cal F}_{m_1, m_2, n_1, n_2} ( 1, 1 ;q) =\begin{cases}
-  2  \frac{\th_2 ^2 \th_4 ^2}{\eta^{4}} \quad   &\text{for } \{m_1, m_2, n_1, n_2\}  \in  \mathbb{Z},  \\
 - 2  (-1)^{m_1} \frac{\th_2 ^2 \th_4 ^2}{\eta^{4}}  \quad   &\text{for }
 \{m_1, m_2, n_2\}  \in  \mathbb{Z}, \;\;  \{n_1\} \in \mathbb{Z } + \frac{1}{2}. 
 \end{cases} 
\end{eqnarray}

\providecommand{\href}[2]{#2}\begingroup\raggedright\endgroup


\begin{thebibliography}{10}

\bibitem{Kachru:1995wm}
S.~Kachru and C.~Vafa, {\it {Exact results for N=2 compactifications of
  heterotic strings}},
  \href{http://dx.doi.org/10.1016/0550-3213(95)00307-E}{{\em Nucl. Phys.}
  {\bfseries B450} (1995) 69--89},
\href{http://arxiv.org/abs/hep-th/9505105}{{\ttfamily arXiv:hep-th/9505105
  [hep-th]}}.

\bibitem{Green:1984bx}
M.~B. Green, J.~H. Schwarz, and P.~C. West, {\it {Anomaly Free Chiral Theories
  in Six-Dimensions}},
\href{http://dx.doi.org/10.1016/0550-3213(85)90222-6}{{\em Nucl. Phys.}
  {\bfseries B254} (1985) 327--348}.

\bibitem{Walton:1987bu}
M.~A. Walton, {\it {The Heterotic String on the Simplest Calabi-yau Manifold
  and Its Orbifold Limits}},
\href{http://dx.doi.org/10.1103/PhysRevD.37.377}{{\em Phys. Rev.} {\bfseries
  D37} (1988) 377}.

\bibitem{Dixon:1990pc}
L.~J. Dixon, V.~Kaplunovsky, and J.~Louis, {\it {Moduli dependence of string
  loop corrections to gauge coupling constants}},
\href{http://dx.doi.org/10.1016/0550-3213(91)90490-O}{{\em Nucl. Phys.}
  {\bfseries B355} (1991) 649--688}.

\bibitem{Mayr:1995rx}
P.~Mayr and S.~Stieberger, {\it {Moduli dependence of one loop gauge couplings
  in (0,2) compactifications}},
  \href{http://dx.doi.org/10.1016/0370-2693(95)00683-C}{{\em Phys. Lett.}
  {\bfseries B355} (1995) 107--116},
\href{http://arxiv.org/abs/hep-th/9504129}{{\ttfamily arXiv:hep-th/9504129
  [hep-th]}}.

\bibitem{deWit:1995zg}
B.~de~Wit, V.~Kaplunovsky, J.~Louis, and D.~L{\"u}st, {\it {Perturbative
  couplings of vector multiplets in N=2 heterotic string vacua}},
  \href{http://dx.doi.org/10.1016/0550-3213(95)00291-Y}{{\em Nucl. Phys.}
  {\bfseries B451} (1995) 53--95},
\href{http://arxiv.org/abs/hep-th/9504006}{{\ttfamily arXiv:hep-th/9504006
  [hep-th]}}.

\bibitem{Harvey-Moore}
J.~A. Harvey and G.~W. Moore, {\it {Algebras, BPS states, and strings}},
  \href{http://dx.doi.org/10.1016/0550-3213(95)00605-2}{{\em Nucl.Phys.}
  {\bfseries B463} (1996) 315--368},
\href{http://arxiv.org/abs/hep-th/9510182}{{\ttfamily arXiv:hep-th/9510182
  [hep-th]}}.

\bibitem{deWit:1996wq}
B.~de~Wit, G.~Lopes~Cardoso, D.~L{\"u}st, T.~Mohaupt, and S.-J. Rey, {\it
  {Higher order gravitational couplings and modular forms in N=2, D = 4
  heterotic string compactifications}},
  \href{http://dx.doi.org/10.1016/S0550-3213(96)90141-8}{{\em Nucl. Phys.}
  {\bfseries B481} (1996) 353--388},
\href{http://arxiv.org/abs/hep-th/9607184}{{\ttfamily arXiv:hep-th/9607184
  [hep-th]}}.

\bibitem{Cardoso-Curio-Lust}
G.~Lopes~Cardoso, G.~Curio, and D.~L{\"u}st, {\it {Perturbative couplings and
  modular forms in N=2 string models with a Wilson line}},
  \href{http://dx.doi.org/10.1016/S0550-3213(97)00047-3}{{\em Nucl.Phys.}
  {\bfseries B491} (1997) 147--183},
\href{http://arxiv.org/abs/hep-th/9608154}{{\ttfamily arXiv:hep-th/9608154
  [hep-th]}}.

\bibitem{Chaudhuri:1995fk}
S.~Chaudhuri, G.~Hockney, and J.~D. Lykken, {\it {Maximally supersymmetric
  string theories in {$D < 10$} }},
  \href{http://dx.doi.org/10.1103/PhysRevLett.75.2264}{{\em Phys. Rev. Lett.}
  {\bfseries 75} (1995) 2264--2267},
\href{http://arxiv.org/abs/hep-th/9505054}{{\ttfamily arXiv:hep-th/9505054
  [hep-th]}}.

\bibitem{Chaudhuri:1995bf}
S.~Chaudhuri and J.~Polchinski, {\it {Moduli space of CHL strings}},
  \href{http://dx.doi.org/10.1103/PhysRevD.52.7168}{{\em Phys. Rev.} {\bfseries
  D52} (1995) 7168--7173},
\href{http://arxiv.org/abs/hep-th/9506048}{{\ttfamily arXiv:hep-th/9506048
  [hep-th]}}.

\bibitem{Aspinwall:1995fw}
P.~S. Aspinwall, {\it {Some relationships between dualities in string theory}},
   \href{http://dx.doi.org/10.1016/0920-5632(96)00004-7}{{\em Nucl. Phys. Proc.
  Suppl.} {\bfseries 46} (1996) 30--38},
\href{http://arxiv.org/abs/hep-th/9508154}{{\ttfamily arXiv:hep-th/9508154
  [hep-th]}}.

\bibitem{Ferrara:1989nm}
S.~Ferrara and C.~Kounnas, {\it {Extended Supersymmetry in Four-dimensional
  Type {II} Strings}},
\href{http://dx.doi.org/10.1016/0550-3213(89)90335-0}{{\em Nucl. Phys.}
  {\bfseries B328} (1989) 406}.

\bibitem{Schwarz:1995bj}
J.~H. Schwarz and A.~Sen, {\it {Type IIA dual of the six-dimensional CHL
  compactification}},
  \href{http://dx.doi.org/10.1016/0370-2693(95)00952-H}{{\em Phys. Lett.}
  {\bfseries B357} (1995) 323--328},
\href{http://arxiv.org/abs/hep-th/9507027}{{\ttfamily arXiv:hep-th/9507027
  [hep-th]}}.

\bibitem{Antoniadis:1992sa}
I.~Antoniadis, E.~Gava, and K.~S. Narain, {\it {Moduli corrections to
  gravitational couplings from string loops}},
  \href{http://dx.doi.org/10.1016/0370-2693(92)90009-S}{{\em Phys. Lett.}
  {\bfseries B283} (1992) 209--212},
\href{http://arxiv.org/abs/hep-th/9203071}{{\ttfamily arXiv:hep-th/9203071
  [hep-th]}}.

\bibitem{Antoniadis:1992rq}
I.~Antoniadis, E.~Gava, and K.~S. Narain, {\it {Moduli corrections to gauge and
  gravitational couplings in four-dimensional superstrings}},
  \href{http://dx.doi.org/10.1016/0550-3213(92)90672-X}{{\em Nucl. Phys.}
  {\bfseries B383} (1992) 93--109},
\href{http://arxiv.org/abs/hep-th/9204030}{{\ttfamily arXiv:hep-th/9204030
  [hep-th]}}.

\bibitem{Cecotti:1992qh}
S.~Cecotti, P.~Fendley, K.~A. Intriligator, and C.~Vafa, {\it {A New
  supersymmetric index}},
  \href{http://dx.doi.org/10.1016/0550-3213(92)90572-S}{{\em Nucl. Phys.}
  {\bfseries B386} (1992) 405--452},
\href{http://arxiv.org/abs/hep-th/9204102}{{\ttfamily arXiv:hep-th/9204102
  [hep-th]}}.

\bibitem{Cecotti:1992vy}
S.~Cecotti and C.~Vafa, {\it {Ising model and N=2 supersymmetric theories}},
  \href{http://dx.doi.org/10.1007/BF02098023}{{\em Commun. Math. Phys.}
  {\bfseries 157} (1993) 139--178},
\href{http://arxiv.org/abs/hep-th/9209085}{{\ttfamily arXiv:hep-th/9209085
  [hep-th]}}.

\bibitem{Kachru}
M.~C. Cheng, X.~Dong, J.~Duncan, J.~Harvey, S.~Kachru, {\em et al.}, {\it
  {Mathieu Moonshine and N=2 String Compactifications}},
  \href{http://dx.doi.org/10.1007/JHEP09(2013)030}{{\em JHEP} {\bfseries 1309}
  (2013) 030},
\href{http://arxiv.org/abs/1306.4981}{{\ttfamily arXiv:1306.4981 [hep-th]}}.

\bibitem{justin1}
J.~R. David, D.~P. Jatkar, and A.~Sen, {\it {Product representation of Dyon
  partition function in CHL models}},
  \href{http://dx.doi.org/10.1088/1126-6708/2006/06/064}{{\em JHEP} {\bfseries
  0606} (2006) 064},
\href{http://arxiv.org/abs/hep-th/0602254}{{\ttfamily arXiv:hep-th/0602254
  [hep-th]}}.

\bibitem{Angelantonj:2011br}
C.~Angelantonj, I.~Florakis, and B.~Pioline, {\it {A new look at one-loop
  integrals in string theory}},
  \href{http://dx.doi.org/10.4310/CNTP.2012.v6.n1.a4}{{\em Commun. Num. Theor.
  Phys.} {\bfseries 6} (2012) 159--201},
\href{http://arxiv.org/abs/1110.5318}{{\ttfamily arXiv:1110.5318 [hep-th]}}.

\bibitem{Angelantonj:2012gw}
C.~Angelantonj, I.~Florakis, and B.~Pioline, {\it {One-Loop BPS amplitudes as
  BPS-state sums}},  \href{http://dx.doi.org/10.1007/JHEP06(2012)070}{{\em
  JHEP} {\bfseries 06} (2012) 070},
\href{http://arxiv.org/abs/1203.0566}{{\ttfamily arXiv:1203.0566 [hep-th]}}.

\bibitem{Angelantonj:2013eja}
C.~Angelantonj, I.~Florakis, and B.~Pioline, {\it {Rankin-Selberg methods for
  closed strings on orbifolds}},
  \href{http://dx.doi.org/10.1007/JHEP07(2013)181}{{\em JHEP} {\bfseries 07}
  (2013) 181},
\href{http://arxiv.org/abs/1304.4271}{{\ttfamily arXiv:1304.4271 [hep-th]}}.

\bibitem{Angelantonj:2015rxa}
C.~Angelantonj, I.~Florakis, and B.~Pioline, {\it {Threshold corrections,
  generalised prepotentials and Eichler integrals}},
  \href{http://dx.doi.org/10.1016/j.nuclphysb.2015.06.009}{{\em Nucl. Phys.}
  {\bfseries B897} (2015) 781--820},
\href{http://arxiv.org/abs/1502.00007}{{\ttfamily arXiv:1502.00007 [hep-th]}}.

\bibitem{Stieberger:1998yi}
S.~Stieberger, {\it {(0,2) heterotic gauge couplings and their M theory
  origin}},  \href{http://dx.doi.org/10.1016/S0550-3213(98)00770-6}{{\em
  Nucl.Phys.} {\bfseries B541} (1999) 109--144},
\href{http://arxiv.org/abs/hep-th/9807124}{{\ttfamily arXiv:hep-th/9807124
  [hep-th]}}.

\bibitem{Dijkgraaf:1996it}
R.~Dijkgraaf, E.~P. Verlinde, and H.~L. Verlinde, {\it {Counting dyons in N=4
  string theory}},  \href{http://dx.doi.org/10.1016/S0550-3213(96)00640-2}{{\em
  Nucl. Phys.} {\bfseries B484} (1997) 543--561},
\href{http://arxiv.org/abs/hep-th/9607026}{{\ttfamily arXiv:hep-th/9607026
  [hep-th]}}.

\bibitem{LopesCardoso:2004xf}
G.~Lopes~Cardoso, B.~de~Wit, J.~Kappeli, and T.~Mohaupt, {\it {Asymptotic
  degeneracy of dyonic N = 4 string states and black hole entropy}},
  \href{http://dx.doi.org/10.1088/1126-6708/2004/12/075}{{\em JHEP} {\bfseries
  12} (2004) 075},
\href{http://arxiv.org/abs/hep-th/0412287}{{\ttfamily arXiv:hep-th/0412287
  [hep-th]}}.

\bibitem{Shih:2005uc}
D.~Shih, A.~Strominger, and X.~Yin, {\it {Recounting Dyons in N=4 string
  theory}},  \href{http://dx.doi.org/10.1088/1126-6708/2006/10/087}{{\em JHEP}
  {\bfseries 10} (2006) 087},
\href{http://arxiv.org/abs/hep-th/0505094}{{\ttfamily arXiv:hep-th/0505094
  [hep-th]}}.

\bibitem{Gaiotto:2005hc}
D.~Gaiotto, {\it {Re-recounting dyons in N=4 string theory}},
\href{http://arxiv.org/abs/hep-th/0506249}{{\ttfamily arXiv:hep-th/0506249
  [hep-th]}}.

\bibitem{Hohenegger:2011us}
S.~Hohenegger and S.~Stieberger, {\it {BPS Saturated String Amplitudes: K3
  Elliptic Genus and Igusa Cusp Form}},
  \href{http://dx.doi.org/10.1016/j.nuclphysb.2011.11.012}{{\em Nucl. Phys.}
  {\bfseries B856} (2012) 413--448},
\href{http://arxiv.org/abs/1108.0323}{{\ttfamily arXiv:1108.0323 [hep-th]}}.

\bibitem{Jatkar:2005bh}
D.~P. Jatkar and A.~Sen, {\it {Dyon spectrum in CHL models}},
  \href{http://dx.doi.org/10.1088/1126-6708/2006/04/018}{{\em JHEP} {\bfseries
  04} (2006) 018},
\href{http://arxiv.org/abs/hep-th/0510147}{{\ttfamily arXiv:hep-th/0510147
  [hep-th]}}.

\bibitem{David:2006yn}
J.~R. David and A.~Sen, {\it {CHL Dyons and Statistical Entropy Function from
  D1-D5 System}},  \href{http://dx.doi.org/10.1088/1126-6708/2006/11/072}{{\em
  JHEP} {\bfseries 11} (2006) 072},
\href{http://arxiv.org/abs/hep-th/0605210}{{\ttfamily arXiv:hep-th/0605210
  [hep-th]}}.

\bibitem{Nikulin}
V.~V. Nikulin, {\it {Finite automorphism groups of K\"{a}hler {$K3$}
  surfaces}},  {\em Trans. Moscow. Math. Soc.} {\bfseries 38} (1979) 71.

\bibitem{Candelas:1985en}
P.~Candelas, G.~T. Horowitz, A.~Strominger, and E.~Witten, {\it {Vacuum
  Configurations for Superstrings}},
\href{http://dx.doi.org/10.1016/0550-3213(85)90602-9}{{\em Nucl. Phys.}
  {\bfseries B258} (1985) 46--74}.

\bibitem{Witten:1985xc}
E.~Witten, {\it {Symmetry Breaking Patterns in Superstring Models}},
\href{http://dx.doi.org/10.1016/0550-3213(85)90603-0}{{\em Nucl. Phys.}
  {\bfseries B258} (1985) 75}.

\bibitem{Eguchi:1980jx}
T.~Eguchi, P.~B. Gilkey, and A.~J. Hanson, {\it {Gravitation, Gauge Theories
  and Differential Geometry}},
\href{http://dx.doi.org/10.1016/0370-1573(80)90130-1}{{\em Phys. Rept.}
  {\bfseries 66} (1980) 213}.

\bibitem{Cheng:2010pq}
M.~C. Cheng, {\it {K3 Surfaces, N=4 Dyons, and the Mathieu Group M24}},
  \href{http://dx.doi.org/10.4310/CNTP.2010.v4.n4.a2}{{\em
  Commun.Num.Theor.Phys.} {\bfseries 4} (2010) 623--658},
\href{http://arxiv.org/abs/1005.5415}{{\ttfamily arXiv:1005.5415 [hep-th]}}.

\bibitem{Gaberdiel:2010ch}
M.~R. Gaberdiel, S.~Hohenegger, and R.~Volpato, {\it {Mathieu twining
  characters for K3}},  \href{http://dx.doi.org/10.1007/JHEP09(2010)058}{{\em
  JHEP} {\bfseries 09} (2010) 058},
\href{http://arxiv.org/abs/1006.0221}{{\ttfamily arXiv:1006.0221 [hep-th]}}.

\bibitem{Gaberdiel:2010ca}
M.~R. Gaberdiel, S.~Hohenegger, and R.~Volpato, {\it {Mathieu Moonshine in the
  elliptic genus of K3}},
  \href{http://dx.doi.org/10.1007/JHEP10(2010)062}{{\em JHEP} {\bfseries 10}
  (2010) 062},
\href{http://arxiv.org/abs/1008.3778}{{\ttfamily arXiv:1008.3778 [hep-th]}}.

\bibitem{Eguchi:2010fg}
T.~Eguchi and K.~Hikami, {\it {Note on Twisted Elliptic Genus of K3 Surface}},
  \href{http://dx.doi.org/10.1016/j.physletb.2010.10.017}{{\em Phys. Lett.}
  {\bfseries B694} (2011) 446--455},
\href{http://arxiv.org/abs/1008.4924}{{\ttfamily arXiv:1008.4924 [hep-th]}}.

\bibitem{Eguchi:1988vra}
T.~Eguchi, H.~Ooguri, A.~Taormina, and S.-K. Yang, {\it {Superconformal
  Algebras and String Compactification on Manifolds with SU(N) Holonomy}},
\href{http://dx.doi.org/10.1016/0550-3213(89)90454-9}{{\em Nucl. Phys.}
  {\bfseries B315} (1989) 193}.

\bibitem{Eguchi:2010ej}
T.~Eguchi, H.~Ooguri, and Y.~Tachikawa, {\it {Notes on the K3 Surface and the
  Mathieu group $M_{24}$}},
  \href{http://dx.doi.org/10.1080/10586458.2011.544585}{{\em Exper. Math.}
  {\bfseries 20} (2011) 91--96},
\href{http://arxiv.org/abs/1004.0956}{{\ttfamily arXiv:1004.0956 [hep-th]}}.

\bibitem{Govindarajan:2009qt}
S.~Govindarajan and K.~Gopala~Krishna, {\it {BKM Lie superalgebras from dyon
  spectra in Z(N) CHL orbifolds for composite N}},
  \href{http://dx.doi.org/10.1007/JHEP05(2010)014}{{\em JHEP} {\bfseries 05}
  (2010) 014},
\href{http://arxiv.org/abs/0907.1410}{{\ttfamily arXiv:0907.1410 [hep-th]}}.

\bibitem{Kawai:1996te}
T.~Kawai, {\it {String duality and modular forms}},
  \href{http://dx.doi.org/10.1016/S0370-2693(97)00146-9}{{\em Phys. Lett.}
  {\bfseries B397} (1997) 51--62},
\href{http://arxiv.org/abs/hep-th/9607078}{{\ttfamily arXiv:hep-th/9607078
  [hep-th]}}.

\bibitem{LopesCardoso:1996zj}
G.~Lopes~Cardoso, {\it {Perturbative gravitational couplings and Siegel modular
  forms in D = 4, N=2 string models}},
  \href{http://dx.doi.org/10.1016/S0920-5632(97)00314-9}{{\em Nucl. Phys. Proc.
  Suppl.} {\bfseries 56B} (1997) 94--101},
\href{http://arxiv.org/abs/hep-th/9612200}{{\ttfamily arXiv:hep-th/9612200
  [hep-th]}}.

\bibitem{Kawai:1995hy}
T.~Kawai, {\it {N=2 heterotic string threshold correction, K3 surface and
  generalized Kac-Moody superalgebra}},
  \href{http://dx.doi.org/10.1016/0370-2693(96)00052-4}{{\em Phys. Lett.}
  {\bfseries B372} (1996) 59--64},
\href{http://arxiv.org/abs/hep-th/9512046}{{\ttfamily arXiv:hep-th/9512046
  [hep-th]}}.

\bibitem{justin2}
J.~R. David, D.~P. Jatkar, and A.~Sen, {\it {Dyon Spectrum in N=4
  Supersymmetric Type II String Theories}},
  \href{http://dx.doi.org/10.1088/1126-6708/2006/11/073}{{\em JHEP} {\bfseries
  0611} (2006) 073},
\href{http://arxiv.org/abs/hep-th/0607155}{{\ttfamily arXiv:hep-th/0607155
  [hep-th]}}.

\bibitem{Angelantonj:2014dia}
C.~Angelantonj, I.~Florakis, and M.~Tsulaia, {\it {Universality of Gauge
  Thresholds in Non-Supersymmetric Heterotic Vacua}},
  \href{http://dx.doi.org/10.1016/j.physletb.2014.08.001}{{\em Phys. Lett.}
  {\bfseries B736} (2014) 365--370},
\href{http://arxiv.org/abs/1407.8023}{{\ttfamily arXiv:1407.8023 [hep-th]}}.

\bibitem{Angelantonj:2015nfa}
C.~Angelantonj, I.~Florakis, and M.~Tsulaia, {\it {Generalised universality of
  gauge thresholds in heterotic vacua with and without supersymmetry}},
  \href{http://dx.doi.org/10.1016/j.nuclphysb.2015.09.007}{{\em Nucl. Phys.}
  {\bfseries B900} (2015) 170--197},
\href{http://arxiv.org/abs/1509.00027}{{\ttfamily arXiv:1509.00027 [hep-th]}}.

\end{thebibliography}
\end{document}